\documentclass[12pt,preprint]{aastex}

\usepackage{epsfig}

\newcommand{\nqagb}{$^{14}$N($\alpha$,$\gamma$)$^{18}$F($\beta^+\nu$)$^{18}$O~($\alpha$,$\gamma$)$^{22}$Ne~}
\newcommand{\msun}{\ensuremath{\, \mem{M}_\odot}}
\newcommand{\n}{\ensuremath{\mem{n}}}
\newcommand{\lei}{\ensuremath{^{1}\mem{L}}}
\newcommand{\p}{\ensuremath{\mem{p}}}

\newcommand{\hevi}{\ensuremath{^{4}\mem{He}}}

\newcommand{\cdr}{\ensuremath{^{13}\mem{C}}}
\newcommand{\czw}{\ensuremath{^{12}\mem{C}}}

\newcommand{\cvi}{\ensuremath{^{14}\mem{C}}}
\newcommand{\cfu}{\ensuremath{^{15}\mem{C}}}
\newcommand{\ndr}{\ensuremath{^{13}\mem{N}}}
\newcommand{\nvi}{\ensuremath{^{14}\mem{N}}}
\newcommand{\nfu}{\ensuremath{^{15}\mem{N}}}

\newcommand{\ose}{\ensuremath{^{16}\mem{O}}}
\newcommand{\osi}{\ensuremath{^{17}\mem{O}}}
\newcommand{\oac}{\ensuremath{^{18}\mem{O}}}

\newcommand{\fne}{\ensuremath{^{19}\mem{F}}}
\newcommand{\nezwa}{\ensuremath{^{20}\mem{Ne}}}
\newcommand{\neei}{\ensuremath{^{21}\mem{Ne}}}
\newcommand{\nezw}{\ensuremath{^{22}\mem{Ne}}}

\newcommand{\nazw}{\ensuremath{^{22}\mem{Na}}}
\newcommand{\nadr}{\ensuremath{^{23}\mem{Na}}}
\newcommand{\mgvi}{\ensuremath{^{24}\mem{Mg}}}
\newcommand{\mgfu}{\ensuremath{^{25}\mem{Mg}}}
\newcommand{\mgse}{\ensuremath{^{26}\mem{Mg}}}

\newcommand{\alse}{\ensuremath{^{26}\mem{Al}}}
\newcommand{\alsi}{\ensuremath{^{27}\mem{Al}}}

\newcommand{\siac}{\ensuremath{^{28}\mem{Si}}}
\newcommand{\sine}{\ensuremath{^{29}\mem{Si}}}
\newcommand{\sidr}{\ensuremath{^{30}\mem{Si}}}

\newcommand{\pei}{\ensuremath{^{31}\mem{P}}}
\newcommand{\szw}{\ensuremath{^{32}\mem{S}}}
\newcommand{\fese}{\ensuremath{^{56}\mem{Fe}}}
\newcommand{\fesi}{\ensuremath{^{57}\mem{Fe}}}
\newcommand{\feac}{\ensuremath{^{58}\mem{Fe}}}
\newcommand{\cone}{\ensuremath{^{59}\mem{Co}}}
\newcommand{\niac}{\ensuremath{^{58}\mem{Ni}}}
\newcommand{\nine}{\ensuremath{^{59}\mem{Ni}}}
\newcommand{\nise}{\ensuremath{^{60}\mem{Ni}}}
\newcommand{\niei}{\ensuremath{^{61}\mem{Ni}}}
\newcommand{\nizw}{\ensuremath{^{62}\mem{Ni}}}

\newcommand{\gdr}{\ensuremath{^{63}\mem{G}}}

\newcommand{\krse}{\ensuremath{^{86}\mem{Kr}}}
\newcommand{\rbsi}{\ensuremath{^{87}\mem{Rb}}}
\newcommand{\zrse}{\ensuremath{^{96}\mem{Zr}}}
\newcommand{\zrvi}{\ensuremath{^{94}\mem{Zr}}}
\newcommand{\srac}{\ensuremath{^{88}\mem{Sr}}}
\newcommand{\baac}{\ensuremath{^{138}\mem{Ba}}}
\newcommand{\blac}{\ensuremath{^{208}\mem{Pb}}}

\newcommand{\mem}[1]{\ensuremath{\mathrm{ #1}}}
\newcommand{\spr}{\mbox{$s$-process}}
\newcommand{\kelv}{\ensuremath{\,\mathrm K}}
\newcommand{\abb}[1]{Figure\,\ref{#1}}
\newcommand{\kap}[1]{\S\,\ref{#1}}
\newcommand{\jahre}{\ensuremath{\, \mathrm{yr}}}
\newcommand{\glt}[1]{Eq.\,(\ref{#1})}
\newcommand{\glp}[1]{(Eq.\,\ref{#1})}
\newcommand{\tab}[1]{Table\,\ref{#1}}

\usepackage{natbib}

\shorttitle{$s$-Process and rotation}
\shortauthors{Herwig \etal.}

\received{2002 November 26}
\begin{document}

\title{The $s$-Process in Rotating AGB Stars} 
\author{Falk Herwig}
\affil{Department of Physics and Astronomy, University of Victoria,
  3800 Finnerty Rd, Victoria, BC, V8P 1A1 Canada} \email{fherwig@uvastro.phys.uvic.ca}
\author{Norbert Langer}
\affil{Astronomical Institute, Universiteit Utrecht, P.O. Box 80000, NL-3508 TA Utrecht, The Netherlands} \email{N.Langer@astro.uu.nl}
\author{Maria Lugaro}
\affil{Institute of Astronomy, University of Cambridge, Madingley Road, Cambridge CB3 0HA, United Kingdom } \email{mal@ast.cam.ac.uk}

\begin{abstract}
We model the nucleosynthesis during the thermal pulse phase of a rotating,
solar metallicity
Asymptotic Giant Branch (AGB) star of $3\msun$, which was evolved from a
main sequence model rotating with $250 \mathrm{km/s}$ at the stellar
equator.
Rotationally induced mixing during the thermal pulses produces a
layer ($\sim 2\cdot 10^{-5}\msun$) on top of the CO-core where large
amounts of  protons and \czw\ co-exist.
With a post-processing  nucleosynthesis and mixing code,
we follow the abundance evolution in
this layer, in particular that of the neutron source \cdr\
and of the neutron poison \nvi.
In our AGB model mixing persists during the entire
interpulse phase due to the steep angular velocity gradient at the
core-envelope interface, thereby spreading \nvi\ over the
entire \cdr-rich part of the layer. We follow the neutron production
during the interpulse phase, and find a
resulting maximum neutron exposure  of
$\tau_\mem{max} =0.04 \mem{mbarn^{-1}}$, which is too small to
produce any significant $s$-process.
In parametric models, we then investigate the combined effects of
diffusive overshooting from the convective envelope and
rotationally induced mixing.
Just adding the overshooting and leaving the rotational mixing
unchanged results also in a small maximum neutron exposure
($0.03\mem{mbarn}^{-1}$).
Models with overshoot and
weaker interpulse mixing --- as perhaps expected from more slowly rotating
stars --- yield larger neutron exposures. A model with overshooting
without any interpulse mixing obtained up to $0.72 \mem{mbarn}^{-1}$,
which is larger than required by observations.
We conclude that the incorporation of rotationally induce mixing processes
has important consequences for the production of heavy elements in AGB
stars.
While through a distribution of initial rotation rates,
it may lead to a natural spread in the neutron exposures
obtained in AGB stars of a given mass in general --- as appears to be
required by
observations --- it may moderate the large neutron exposures found in
models with diffusive overshoot in particular.
Our results suggest that both processes, diffusive overshoot and
rotational
mixing, may be required to obtain a consistent description of the
\spr\ in AGB stars which fulfils all observational constraints.
Finally, we find
that mixing due to rotation within our
current framework does increase the production of $^{15}$N in the
partial mixing zone. However, this increase is not large enough to
boost the production of fluorine to the level required by observations.
\end{abstract}

\keywords{ stars: AGB and post-AGB
--- stars: evolution
--- nuclear reactions, nucleosynthesis, abundances
--- stars: rotation
--- stars: interiors}

\section{Introduction}
\label{sec:intro}

Trans-iron elements are mainly made by neutron-capture reactions on \fese\
seed nuclei. Two processes have been been distinguished according to the
neutron density at the production site. In the case of the $r$-process
the n-densities are high ($\mem{N_n} > 10^{20}
\mem{cm}^{-3}$), and  the time scale of successive n-capture reactions
on heavy isotopes is faster than the $\beta$-decay time scale. Such a
sudden high-density neutron burst creates isotopes far away from the
valley
of $\beta$-stability in the chart of nuclides, which successively decay back
to the stable isotopes. In contrast the \spr\ is characterized by
lower neutron densities ($\mem{N_n} \lessapprox 10^{10}
\mem{cm}^{-3}$). Neutron captures are generally followed by $\beta$-decays
since unstable isotopes on the \spr\ path have typical life times of the
order of hours. In some cases, however, the unstable isotopes
involved have longer life times and, depending on the neutron density
and temperature conditions,
$branchings$ can be open on the \spr\ path leading to the production of
neutron-rich isotopes \citep[see][ for an introduction to the
\spr]{clayton:68}. 

In Asymptotic Giant Branch (AGB) stars recurrent He-shell flashes
(thermal pulses, TP) drive a convective zone that temporarily covers
the whole 
region between the H-burning and the He-burning shells (intershell).
Here, partial He burning produces a high mass fraction ($>0.25$) of
\czw\ and the chain \nqagb starting on the abundant \nvi\ from the
H-burning ashes produces a relatively large amount of \nezw\ (mass
fraction $\simeq$ 0.02). The $\nezw(\alpha,n)\mgfu$ reaction as neutron
source for the \spr\ was suggested by \citet{cameron:60}. Temperatures
above 3$\cdot 10^{8}$ K are required for that reaction to release
a significant amount of neutrons.
Stellar models showed that such high temperatures are achieved in
intermediate-mass ($M_\mem{ZAMS}/\msun > 4$) AGB stars, which were hence
proposed as the main site for the production of \spr\ elements
belonging to the solar main component, i.e. 90 $< A <$204
\citep{iben:75,truran:77}. However, the neutron density produced by
$^{22}$Ne burning in thermal pulses is rather high (above
$10^{11}\mem{cm}^{-3}$ for $T = 3.5 \cdot 10^8 \kelv$). This leads to
excesses in the neutron-rich nuclides produced by branchings, for
example \krse, \rbsi, and \zrse\ \citep[see e.g.][]{despain:80}, in
contrast with the great majority of the observations of \spr-enhanced
stars such as MS, S and C stars.  In S and C stars the Rb/Sr ratio is
typically much lower than solar \citep{lambert:95,abia:01}, indicating a
low neutron density at the \spr\ site. Also the Rb abundances observed
in 10 AGB members of the massive Galactic globular cluster $\omega$
Centauri indicate a low neutron density for the \spr\ \citep{smith:00}.
\citet{lambert:95} reported the zirconium isotopic abundance obtained by
spectroscopic observations of the ZrO bandheads in M, MS and S stars
and found no evidence of
an excess of the neutron-rich \zrse\, which can be produced in great
amount by the \spr\ when the neutron density exceeds $\sim 5 \cdot
10^{8}$ $n$ cm$^{-3}$.  Another problem is that the activation of the
$\nezw(\alpha,n)\mgfu$ is expected to produce an excess of $^{25}$Mg in
stars enriched in \spr\ elements. Instead these stars typically have
magnesium isotopic abundances in solar proportion \citep[see
e.g.][]{smith:86,mcwilliam:88}.  Also other types of observations tend
to exclude AGB stars of intermediate mass as the main s-process site. 
Observations mainly show that MS, S and C stars have low luminosity
\citep{frogel:90} and hence low mass. \citet{feast:89} performed a study
of the kinematics of peculiar red giants including S, SC, and C stars.
On the basis of 183 S - SC stars, and 463 C stars he estimated their
mean mass to be 1.3 and 1.6 $\msun$ respectively, although this estimate
needs to be improved.  In summary, the observational evidence and  the
current state of AGB evolution models suggest that
the major nuclear production site of the  \spr\  are low-mass AGB
stars.

In low-mass AGB stars the temperature in the
intershell is not high enough to burn a significant amount of \nezw.
The $\cdr(\alpha,n)\ose$ neutron source reaction, which was suggested by
\citet{greenstein:54} and \citet{cameron:55} and is activated at lower
temperatures ($\sim 0.9 \cdot 10^{8}$ K), is expected to play the major
role \citep{iben:82,gallino:88,kaeppeler:90}.  However, an amount of
\cdr\ higher than that present in the H-burning ashes is needed to
reproduce the observed enhancements of heavy elements. In order to form
a sufficient amount of \cdr\ it is hence speculated that some protons
mix into the \czw-rich intershell \citep[see][ for a general review of
the \spr\ in AGB stars]{busso:99}. 

In recent years a picture of the \spr\ based on these results has emerged
\citep{gallino:98} and is summarized in \abb{fig:schematic}. The He-flash
convection zone homogenizes the intershell region and \czw\ produced in the
He-burning shell is mixed up to just below the location of the 
H-burning shell. The dashed line in \abb{fig:schematic} indicates that,
after the convective pulse is extinguished, the convectively unstable
envelope may extend down into H-free layers of the intershell region.
This phenomenon allows processed intershell material to be carried into
the envelope and hence to the stellar surface 
({\it third dredge-up}). At the end of third dredge-up a layer is
created where the H-rich envelope directly neighbours the \czw-rich
intershell. This layer is a favourable region for the formation of 
a zone where \czw\ and protons are partially mixed. As the temperature
increases in the region a pocket of \cdr\ forms by proton captures on \czw. 
Subsequently, the \cdr\
serves as a neutron donor via the reaction $\cdr(\alpha,\n)\ose$ which is
activated during the following interpulse phase at $T\sim 
9 \cdot 10^\mem{7}\kelv$ so that neutrons are released under radiative
conditions \citep{straniero:95}. Typically the \spr\ occurs on a time scale of
several tens of thousands years and before the onset of the following
thermal pulse. In the convective pulse the $^{22}$Ne neutron source is
only marginally activated.

It is most reasonable to assume that the H/\czw-ratio in the partial mixing
zone
varies somehow continuously from a few hundreds in the envelope to zero
in the intershell. Then,
one finds in the top layers of the partial mixing zone another pocket
made of \nvi, which forms where the H/\czw-ratio is larger than where the
\cdr\ pocket forms. Without further mixing the partial mixing layer is
strictly stratified during the interpulse period: as shown in
\abb{fig:schematic}
the two pockets coexist without much interaction. During the interpulse
phase the temperature does not reach values required for
the $\nvi(\alpha,\gamma)^{18}$F reaction. The \nvi\ pocket is engulfed into the
following He-flash convection zone where it might slightly contribute to the
production of $\nezw$.

Stellar models that use a standard treatment of convective mixing can not
produce the \cdr\ pocket because extra mixing processes are required to
allow a small amount of protons to enter the \czw-rich region. \citet[][ 
and following studies by that group]{gallino:98} and \citet{goriely:00}
have studied the \spr\ by assuming a certain proton profile extending
into the
\czw-rich region without relating explicitly to a specific physical
process. In these studies it was  implicitly assumed that any
subsequent alterations of
the abundances in the partial mixing zone are due to nucleosynthesis only and
no mixing takes place during the interpulse phase after the initial formation
of the partial mixing zone. In this way it was possible to develop a fairly
good understanding of the \spr\ nucleosynthesis in the \cdr\ pocket. For
example, it was shown that many observations are reproduced with models in which the neutron exposure  in the \cdr\ pocket is
up to $\simeq 0.4\mem{mbarn}^{-1}$ at  solar metallicity. In order to
account for the observed \spr\ overabundances the partial mixing zone
needs to have a mass of 
$10^{-4}$ - $10^{-3} \msun$. These \spr\ models with an assumed
H-profile for the partial mixing zone can account for many of the
overall observed properties of the \spr. 

However, to explain the observed \spr\ signature in AGB stars as a 
function of
metallicity, the \spr\ model described above requires that stars of the same
mass and metallicity have different neutron exposures due to the \cdr\ neutron
source during the AGB interpulse phase \citep{busso:01a}.
A mixing process for the formation and evolution of this neutron
source that allows for some \emph{spread in the efficiency of producing
neutrons} seems to be necessary.
\citet{vanwinckel:00} arrive at the same conclusion from
observations of weakly metal-poor post-AGB stars with $21\mem{\mu m}$
feature. In these objects the \spr\ element signatures are easier to measure
than in the cool progenitor AGB stars. 
Also the spread in the Pb abundance observed in very low-metallicity
stars \citep[for example][]{vaneck:03} could fit into the current \spr\ model only if 
there is a spread in the number of neutrons available in the pocket for the
\spr. Such a spread in the efficiency of the \cdr\ neutron source is also
required to explain the measurements of isotopic ratios in single
pre-solar silicon carbide grains (SiC) recovered from pristine
carbonaceous meteorites \citep{zinner:98}. The majority of these grains
are believed to have formed in the circumstellar dusty envelopes
surrounding carbon stars. The main evidence
for this comes from measurements of aggregates of SiC that have shown a very
strong signature of the \spr\ in the heavy elements present in trace
\citep[see e.g.][]{gallino:97a}. 
\citet{nicolussi:97,nicolussi:98a,nicolussi:98b} and \citet{savina:03}
performed measurements of the composition of Sr, Zr, Mo and Ba in single
SiC grains.
Isotopic ratios that are sensitive to the efficiency of the \cdr\ neutron
source have been found to show a large spread within the single grain
data. This can be explained if SiC grains were formed in a multiplicity of
low mass AGB stars with a range of neutron exposures in the \cdr\ pocket.
These isotopic ratios are those involving nuclei with very low neutron capture
cross section such as nuclei with number of neutrons equal or near to nuclear
magic numbers, i.e. $^{88}$Sr/$^{86}$Sr, $^{90,91,92}$Zr/$^{94}$Zr and 
$^{138}$Ba/$^{136}$Ba. A detailed analysis of SiC grain data 
indicated that a spread of the order 5 is necessary in the neutron
exposure for a given mass and metallicity \citep{lugaro:03b}.
These new insights are specifically important for constraining the physical
processes which are responsible for the partial mixing between envelope and
core, and thereby lead to the formation of the \cdr\ neutron source. 
Moreover, neutron capture elements are in general becoming an increasingly
important target of stellar observations, for example of
IR-observations in planetary nebulae \citep{dinerstein:01}.

The open problem addressed in this paper is the role of rotational mixing. 
Rotation is an effect that has to be taken into account when studying
stellar evolution and nucleosynthesis. Most F stars, which are the
main sequence progenitors of low-mass AGB stars, show rotational
velocities of a few hundred $\mem{km}\,\mem{s}^{-1}$ 
\citep{royer:02,royer:03}. 
The importance of rotation as a physical process in AGB stars is not
restricted to mixing and nucleosynthesis. Rotation in stars during their late
evolutionary phase possibly drives the shaping of bipolar proto-planetary
nebulae. According to the interacting-stellar-winds model
\citep{kwok:82,balick:87} a fast ionized stellar wind interacts  with an
equatorially dense AGB circumstellar envelope that could be the result 
of inhomogeneities associated with rotation. This model   
can qualitatively explain the presence of sharp radial structures and the wide
variety of shapes found in planetary nebulae \citep[see
also][]{icke:92}.
\citet{reimers:00}  find that elliptical or weakly bipolar planetary
nebulae shapes
can result from dust-driven winds of rotating AGB stars. \citet{soker:01} 
considers the possibility that increased rotation in an AGB envelope may
result from swallowing another celestial body, 
like a companion star in a binary system or an orbiting planet. 
Indeed,  this hypothesis gains support from the 
recent discovery of many extra-solar planets and the
detection of water vapour around evolved
AGB stars, possibly due to the presence of comets \citep{melnick:01}.
\citet{garcia-segura:99} present hydrodynamical models in which an
equatorial density enhancement originates around an  intermediate mass
single AGB star with a fast rotating core 
that can spin up the extended envelope 
during mixing events associated with the He-shell flashes.

\citet{langer:99} evolved a $3\msun$ stellar model from
the main sequence to the AGB phase including the effects of rotation
on the stellar structure and mixing. They found that
rotationally induced mixing at the core-envelope interface after a thermal
pulse could be responsible for the formation of the
partial mixing zone that hosts a \cdr\ pocket and subsequent \spr\ 
nucleosynthesis. They also found that mixing in the partial mixing layer 
continues throughout the entire interpulse phase.
In this paper we investigate mixing and the 
\spr\ at the core-envelope interface of this stellar model with
rotation. We will also present a comparison with models including
mixing due to hydrodynamic overshooting as well as parametric models
that further illustrate the effect of slow mixing of the \spr\ layer
during the interpulse period.

In the current model the \spr\ occurs in every interpulse-pulse
cycle from when third dredge-up starts until the end of the AGB
evolution. Several stellar parameters that are important for the
computation of the \spr, e.g. the mass of the
intershell, the temperature at the base of the convective shell, the
overlapping  factor between subsequent pulses and the third dredge-up,
are different at each interpulse-pulse cycle.
Detailed \spr\ calculations, such as those of \citet{gallino:98}, 
take into account these effects. However, the features of the \cdr\ pocket 
are kept
the same in each interpulse, and the changes in the temperature are not
large enough to affect the modality of the burning of \cdr\ in
different interpulses. The only effect is that in detailed
calculations the neutron exposure in the pocket slightly decreases with
the interpulse number because the amount of \spr\
material increases in the intershell. 
We perform simulations of the \spr\ over only one interpulse period, which
in first approximation represents all the interpulse periods.
In \kap{sec:obs-pm} we derive the basic properties of the partial
mixing zone and interpulse \spr\ layer from average observational
features and simplified AGB evolution properties. 
We describe the stellar models and the nucleosynthesis code in
\kap{sec:models}. The following section is devoted to our scheme
for the heavy \spr\ neutron sink (\kap{sec:sink}). 
The properties and the effects of mixing induced by hydrodynamic
overshoot and by rotation are presented in \kap{sec:ov} and 
\kap{sec:rotmix} respectively. Mixing for the \spr\ is further explored
with synthetic models in \kap{sec:artmix}. The particular problem of
the production of \fne\ is addressed in \kap{sec:f19} and we present a
final discussion in \kap{sec:disc}. 

\section{Constraints on the Properties of the Partial
Mixing Zone}
\label{sec:obs-pm}

The properties of a partial mixing zone that reproduces the \spr\
features observed in AGB stars can be studied in  detail with models
including the effect of many consecutive thermal pulses and
neutron exposure events \citep{gallino:98,goriely:00}. Here we derive
some basic constraints on the properties of the \spr\ zone by
following a much simpler approach. We consider two \spr\ indicators:  
the index $s/s_{\odot}$ is the overproduction factor of \spr\ elements with
respect to the initial solar value. We have used for this index the average of the
production factors of Y and Nd. The index [hs/ls] = [hs/Fe] - [ls/Fe] monitors
the distribution of the \spr\ elements. We have used 
[ls/Fe]=$\frac{1}{2}$([Y/Fe]+[Zr/Fe]) and
[hs/Fe]=$\frac{1}{5}$([Ba/Fe]+[La/Fe]+[Ce/Fe]+[Nd/Fe]+[Sm/Fe]), where square
brackets indicate the logarithmic ratio with respect to the solar ratio. 
Observationally, the spectroscopic 
studies of the \spr\ abundances in evolved low-mass stars of solar
metallicity can be summarized as $0 < \log(s_\mem{obs}/s_{\odot}) < 1$ and $-0.5 <
\mem{[hs/ls]} < 0$ \citep{busso:95}. 

The observed overproduction factors in the envelope are related to the
overproduction factors in the \spr\ zone by two dilution
factors that result from two subsequent mixing events: the He-flash
convective mixing and the third dredge-up. Assuming that no significant
amount of \spr\ material is
available in the envelope initially and considering only the
contribution of the \spr\ in the interpulse, the abundance of
any species in
the envelope after third dredge-up events in $m$ identical TP 
cycles is
related to the abundance in the \spr\ zone (partial mixing zone, PM) by
\begin{equation}
\label{math:mix}
Y_\mem{env}=q\, m\, Y_\mem{PM}\,\frac{M_\mem{PM}  M_\mem{DUP}}{ M_\mem{IS}
M_\mem{env}}
\end{equation}
where  $M_\mem{DUP}$, $M_\mem{IS}$
and  $M_\mem{env}$ are the masses of the
dredged-up layer, the intershell zone covered by the He-flash
convection and the envelope respectively. In low mass AGB stars with
core masses of $\sim 0.6\msun$ these quantities are of the order  $M_\mem{DUP} 
\sim 3 \cdot 10^{-3}\msun$, $ M_\mem{IS}\sim 10^{-2}\msun$
and  $ M_\mem{env}\sim 0.5\msun$. These masses vary from pulse to
pulse and are dependent on the core mass and on the treatment of
mixing. For example in the model 
sequence with rotation  $ M_\mem{IS} =1 \cdot 10^{-2}\msun$ at a core
mass of $M_\mem{c}=0.746\msun$ \citep{langer:99} while in the sequence
with overshooting at all convective boundaries $ M_\mem{IS} =2.4 \cdot
10^{-2}\msun$ at a core mass of $M_\mem{c}=0.628\msun$ \citep{herwig:99a}.
The mass $M_\mem{PM}$ refers to the layer of the partial
mixing zone that at the end of the interpulse phase contains
\spr\ material. In models without mixing during the
interpulse phase this corresponds to the region of the partial mixing
zone were initially $-3 < \log X(H) < -2$ \citep{goriely:00}. In
models with mixing during the interpulse like in the case with
rotation the extent of the partial mixing zone can only be determined
at the end of the interpulse when the calculation has yielded the
s-process nucleosynthesis and mixing result.
 The factor $q$ describes the effect
of overlapping He-flash convection zones in subsequent TP
(see below). 

Without resorting to the detailed results of full stellar evolution
calculations we estimate the number $m$ of thermal pulses with dredge-up
events that enrich the envelope in a semi-empirical way.
As we have discussed in the introduction carbon stars are
believed to be the result of recurrent third dredge-up events in low
mass stars with initial ZAMS masses predominantly in the range $1.5 < M/\msun 
< 3$ \citep[see also][]{groenewegen:95}. These stars
end their lives as white dwarfs with masses in the range $0.57 < M/\msun <
0.68$, according to the revised stellar initial-final mass relation 
of \citet{weidemann:00}. In fact, the mass distribution of 
white dwarfs peaks at or just below $0.60\msun$
\citep{koester:79,weidemann:84,bergeron:92,napiwotzki:99a}. This mass
distribution is very similar to that of central stars of planetary
nebulae \citep{stasinska:97}, which are in the evolution phase between
AGB stars and white  dwarfs. This means  
that the majority of carbon stars must have achieved the necessary
abundance enrichment through the third dredge-up before or when reaching a
core mass of $\sim 0.6\msun$. According to the synthetic AGB models of
\citet{marigo:96} significant dredge-up must commence at core masses
of $0.58\msun$ or even below in some cases, in order to reproduce the
carbon star luminosity function \citep{marigo:99}.
Therefore the relevant chemical enrichment of
AGB stars typically occurs within an effective core mass growth of
about $\Delta M_\mem{cg} = 0.02\msun$, and maybe up to 0.06\msun\ in
some cases. For low 
mass TP-AGB stars the core mass growth per TP is about $\Delta
M_\mem{H}=6\cdot 10^{-3}\msun$ and with a dredge-up parameter of
$\lambda=0.5$ about m=7 ($\Delta M_\mem{cg}=0.02$) and 
possibly up to 20 thermal pulses with dredge-up mixing (if dredge-up
starts at $M_\mem{c}=0.54\msun$, $\Delta
M_\mem{cg}=0.06$) can be considered to be responsible for the
abundance enrichment.    

Expression \glt{math:mix} does not take into account that processed
heavy elements accumulate in the intershell from one TP to the next
because the He-flash convection zone is partly overlapping (overlap
factor $r$) with layers that have 
been swept by the previous He-flash convection zone. In the case of
nucleosynthesis of a species ($s$) during the interpulse phase in a
partial mixing zone  the production stays roughly constant from TP to TP.
The abundance due to such a production is
$X_\mem{s}=X_\mem{PM}
M_\mem{PM} /  M_\mem{IS}$. The total 
intershell abundance  at the n$^\mem{th}$ TP with third dredge-up is then
given by $X_n=X_\mem{s} + rX_{n-1}$. The number $l$ of TPs
needed to approach a $90\%$ 
level of some asymptotic value for the intershell abundance of species
s is given by  $l=-1/\log r$. For  $n>l$ it follows that 
\begin{equation}
\label{gl:q}
X_n \approx X_\mem{s} \cdot q = X_\mem{s} \cdot \sum_{i=0}^{l-1} r^i 
\end{equation}
The overlap factor decreases as a function of TP number and also
depends on the details of the third dredge-up efficiency. Typically 
overlap factors  
in stellar models decrease from about $0.8$ at the earliest TPs to
an asymptotic value larger than $0.4$. The condition $n>l$ is
approximately  satisfied for $r \lessapprox 0.6$, for which \glt{gl:q} returns
$q=2.3$. Other triplets of $(r,l,q)$ are $(0.4,3,1.6)$ and
$(0.8,11,4.6)$. By evaluating \glt{math:mix} for the numbers specified
above, and with $Y_\mem{env}$ in \glt{math:mix} given by using 
the maximum $\log(s_\mem{obs}/s_{\odot}) = 1$, we  derive a
logarithmic expression that relates the \spr\
overabundance in the PM zone with the mass of that zone:
\begin{equation}
\label{gl:sFe-Mpm}
\log M_\mem{PM}= - log(s_\mem{PM}/s_{\odot}) + c
\end{equation}
where $c=-0.14$ $(0, -0.44)$ for $m=10$ $(7, 20)$.

In \abb{fig:tau-shsls} we show the variation of log($s_\mem{PM}/s_{\odot}$) and
$\mem{[hs/ls]_{PM}}$ with the neutron exposure.
These trends have been obtained by fully implicit network
calculations containing the s-process nucleosynthesis with neutron
captures on all isotopes from He to Pb 
\citep[as in][]{lugaro:03a}, and all relevant charged particle reactions (see
\kap{sec:models} for reaction rate references). As  
more neutrons are released the average \spr\ overabundance 
increases.  In the current \spr\ model the partial mixing zone by definition
does not extend into the He-burning shell, and therefore the partial
mixing zone can not exceed the mass of the intershell layer. Thus, the
mass available for 
the pocket is $M_\mem{PM}< 10^{-2} \msun$. This, together with
\glt{gl:sFe-Mpm} and $m=10$ requires that
$\log(s_\mem{PM}/s_{\odot})>1.86$ and translates into a minimum for
$\tau$ in the 
partial mixing zone. In \abb{fig:tau-shsls} the left shaded part of the
diagram with $\tau < 0.2 \mem{mbarn}^{-1}$  is thereby excluded as the 
predominant region of $\tau$ in the partial mixing zone.  Values of 
$\log(s_\mem{PM}/s_{\odot})<1.86$ 
in that region would require a prohibitively large partial mixing zone
($M_\mem{PM} > 10^{-2} \msun$) in order to reproduce the most \spr\
enriched stars of solar metallicity with $\log(s_\mem{obs}/s_{\odot})
=1$. From [hs/ls]$_\mem{obs}<0$ the
$\tau$-values  
corresponding to the right shaded area in \abb{fig:tau-shsls} can be
excluded as typical for the \spr. In fact, if the neutron exposure
values predominant in the pocket exceed $\sim 0.5$ mbarn$^{-1}$ 
the envelope \spr\ abundance will show [hs/ls] $>$ 0. This
case is discussed in \kap{sec:pp-ov} with an example in
\tab{tab:pocket-mass-test} \citep[see also][]{lugaro:03a}. 
To summarize, we estimate the following properties for the partial
mixing zone in stars of solar metallicity: 
\begin{itemize}
\item the \cdr\ pocket should generate a  predominant neutron
exposure in the range $\sim$ 0.2 - 0.5 mbarn$^{-1}$;
\item using \glt{gl:sFe-Mpm} and $\log(s_\mem{PM}/s_{\odot}) = 4$,
corresponding 
to the upper limit of $\tau$, the mass of the partial mixing zone should obey
$M_\mem{PM} > 7\cdot 10^{-5}\msun$ (assuming $m=10$), in agreement with
the detailed studies. 
\end{itemize}

\section{Stellar Evolution and Nucleosynthesis Models}
\label{sec:models}

Stellar evolution models with rotation are computed in the same way as in 
\citet{langer:99}. The 1D-hydrodynamical stellar evolution code
\citep{langer:98a,heger:00} considers angular momentum, the
effect of centrifugal force on the stellar structure and the following
rotationally induced transport mechanisms for angular momentum and
chemical species: Eddington-Sweet circulation, Solberg-H{\o}iland and
Goldreich-Schubert-Fricke instability, and dynamical and secular shear
instability. The $\mu$-gradient, which acts as a barrier for
rotationally induced mixing, and the Ledoux criterion for
convection and semi-convection are considered. The nuclear energy
generation is computed in the operator split approximation with a
network including  pp-chain, H-burning cycles and He-burning.
For comparison we analyze stellar models with
hydrodynamic overshoot of \citet{herwig:99a}. These models feature a
partial mixing zone of protons and \czw\ very similar to that
assumed for \spr\ calculations with a parameterized
partial mixing zone, like in \citet{goriely:00}. They assume however
that overshoot is to some extent present at all convective boundaries.
The two processes of rotation and overshoot are considered independently
from each other as the
stellar evolution models with rotation do not include
overshoot and the overshoot models do not include rotation. 

The nuclear reactions under consideration for the \spr\ do not
contribute in a
significant way to the energy generation in the star. Hence post-processing is a
valid approximation and a faster and easier-to-handle alternative to
re-computing the whole stellar evolution including all the needed
nuclear species.
An important feature of our rotating AGB stellar models is a weak and 
persistent mixing of the partial 
mixing layer where the \cdr\ and \nvi\ pocket are located next to each
other.  The \spr\ can not be computed under the assumption of
stratification anymore, as done for example by \citet{lugaro:03a}.  
Therefore we have developed a nucleosynthesis and mixing post-processing
code (SBM6) which solves simultaneously 
for mixing processes, charged particle nuclear reactions as well as
neutron production and destruction reactions and $\beta$-decays.  The code
uses thermodynamic and mixing properties of the stellar evolution
models as input and follows the abundance evolution of the 
nuclear network described below.
In the stellar evolution code with rotation  
the time and spatial resolution is determined by the H-burning
shell. Grid rezoning after the
thermal pulse at the core-envelope interface may introduce some
numerical diffusion which can be avoided if the s-process simulation
is carried out on a fixed grid beginning right after the formation of
the partial mixing zone. 
In the post-processing we use a high-resolution, equidistant Lagrangian grid with
400 points  to cover our partial mixing layer which has a
mass of $10^{-4}\msun$. 
The solution scheme of coupled
burning and mixing is fully implicit \citep{herwig:00a} with adaptive
time steps. A Newton-Raphson iteration is accepted as solution if the
corresponding greatest relative correction is less than $10^{-3}$ for
any species at any grid point with a mass 
fraction $X>10^{-21}$. This guarantees a numerical precision of $0.1\%$
for any species at each time step and the correct integration of the
neutron abundance, which is treated like that of any other species. The 
neutron density typically encountered during the interpulse ($N_\mem{n}
< 10^{7} n/\mem{cm^{-3}}$) corresponds to $X(\n) < 2 \cdot 10^{-21}$. In
practise the precision for neutrons is much better than $0.1\%$ even for
$X(\n) < 10^{-21}$ because the
convergence of the set of equation is determined by other species.
Due to the fully implicit solution scheme only a fraction (about one hundredth)
of the time steps of the evolutionary code need to be computed with the SBM6
code because in the partial mixing zone the thermodynamic
conditions change much slower than at the location of the H-burning shell.


For the post-processing simulations we have considered the following reactions in
the SBM6 code:
\begin{itemize}
\item $(\p,\gamma)$ from NACRE compilation \citep{angulo:99} on 
\czw, \cdr, \ndr,\nvi, \nfu, \ose, \osi, \oac, \fne, \nezwa, \neei,
\nezw, \nazw, \nadr, \mgvi, \mgfu  (both to $\alse_\mem{g}$ and
$\alse_\mem{m}$), \mgse, \alsi, \siac,  
\sine\ and  \sidr. \pei\ and $\alse_\mem{g}(\p,\gamma)$ from
\citet{iliadis:01}. $\cvi(\p,\gamma)$ from \citet{wiescher:90}. 
\item $(\p,\alpha)$ from NACRE on \nfu, \osi, \oac\ and \fne.
\item $\beta$-decays have been assumed to follow
instantaneously were applicable, except for 
the $\beta$-decays of \nazw, $\alse_\mem{g}$, \ndr, \cvi\ and \nine\ 
from the Karlsruher Chart of Nuclides.
\item $(\alpha,\gamma)$ from NACRE on \czw, \nfu\ and \oac, and on
  \cvi\ from \citet[][]{gai:87,hashimoto:86,funk:89}\footnote{We have
used the electronic NETGEN database to retrieve these and other rates \citep{jorissen:01}.}.
\item $(\alpha,\n)$ from NACRE on \cdr\ and \nezw.
\item $(\n,\gamma)$ from  the compilation by \citet[][Appendix
B]{lugaro:01a} which is largely based on \citet{bao:00} on
\czw, \nvi, \ose, \neei, \nezwa,  \nezw, \nadr, \mgvi, \mgfu, \mgse,
\alsi, \siac, \sine, \sidr, \pei, \fese, \fesi, \feac, \cone, \niac,
\nine, \nise, \niei\ and
\nizw. 
\item $(\n,\p)$ neutron recycling reactions on $\nvi$ \citep{gledenov:95}
  and $\nine$ from the Reaclib Data Tables of nuclear reaction rates (1991),
updated version of the compilation by \citet{thielemann:87}.
\item A light and a heavy neutron sink accounts for neutron absorbing
species between \szw\ and \fese\ and above \nizw\ respectively (see
\kap{sec:sink} for details). 
\end{itemize}
We consider the light
n-capture reactions that are important to determine the
n-density. The efficiency of isotopes to absorb neutrons can be
measured by the product of the n-capture Maxwellian-averaged cross
section and the 
number density of the isotope. Assuming a solar abundance distribution the
12 most important light n-absorbing isotopes are (in the order of
decreasing absorbing efficiency, $(\n,\gamma)$ reaction unless noted):
 \nvi(\n,\p), \sidr, \alsi, \siac, \sine, \mgfu, \ose, \nvi, \mgvi,
 \nadr, \nezwa\ and \czw. The order of this list of  n-capture
 isotopes changes at locations in stars where a non-solar abundance
 distribution has been established. E.g.\ in the \nvi\ pocket (see
 \kap{sec:ov}) the \nvi\ abundance is so high that the n-captures by
 $\nvi(\n,\gamma)$ reactions outnumber all other $(\n,\gamma)$
reactions on light elements and is itself dwarfed by
 $\nvi(\n,\p)\cvi$. The long-lived ground state of the unstable
 nucleus \alse\ is also a major neutron poison due to its very large
cross sections for $(\n,\p)$ and $(\n,\alpha)$ reactions
\citep{skelton:87,koehler:97}. This nucleus is produced by proton captures on
 \mgfu\  during H-burning, however it is efficiently destroyed by
neutron captures in the thermal pulse convective regions, hence it is
not present in the \cdr\ pocket and we have omitted it in the
computation.

The neutron recycling induced by
the \nvi(\n,\p)\cvi\ reaction is of particular importance for the
\spr\ in rotating AGB stars. This reaction deprives the Fe-seed and
trans-iron elements of available neutrons. The major effect is 
that the neutron exposure is smaller if \nvi\ is present (see Appendix
A). In addition the \nvi(\n,\p)\cvi\ reaction  opens a channel to make
the isotope \nfu, hence potentially \fne. \cvi\ has a half life of
$5730\jahre$. Under the typical conditions of n-production in the
interpulse phase ($T=9 \cdot 10^7\kelv, \rho = 3700 \mem{g cm^{-3}}$)
\cvi\ can capture an $\alpha$ particle or a proton. The
$\cvi(\alpha,\gamma)\oac$ reaction is 2.5 times slower than
$\cdr(\alpha,\n)\ose$,  but still affected
by large uncertainties. The p-capture reaction of \cvi\
($\cvi(\p,\gamma)\nfu$) is 3.7 times faster than
$\czw(\p,\gamma)\ndr$ and about as fast as $\cdr(\p,\gamma)\nvi$.
No other p- or $\alpha$-capture reactions of
\cvi\ are important and also  the $\cvi(\n,\gamma)\cfu$ can be neglected.

\section{Approximating the $s$-Process with a Neutron Sink}
\label{sec:sink}

We introduce two reactions and two artificial particles that 
 together give an estimate  on the overproduction of \spr\ elements
and the \spr\ distribution. First, we consider the reaction
$\nizw(\n,\gamma)\gdr$, where \gdr\ is an artificial particle with
 mass number $63$. We identify the number abundance of \gdr\ with the
combined  number abundance of all isotopes heavier than \nizw:
\begin{equation} 
Y(\gdr)=\sum_{A=63}^{209} Y(^{A}\mem{S}),
\end{equation} 
where S stands for the element symbol of the  respective
species with $A>62$. For a solar composition the number abundance
of \gdr\ is 
$Y_\odot(\gdr)=5.4 \cdot 10^{-8}$. Starting on \fese\ a suitable
neutron exposure will quickly lead to the formation of heavy particles
$^\mem{A}$S (with $A>62$), hence the abundance of  \gdr\ increases. In order  
to count the number of neutron captures that occur on
species with $A>62$ we introduce a second reaction
$\gdr(\n,\lei)\gdr$, which plays the role of heavy neutron sink. The
Maxwellian-averaged cross section of \gdr\ is computed in the usual way
\citep{jorissen:89}:
\begin{equation}
\label{eq:sink-cross}
\sigma(\gdr) =Y(\gdr)^{-1} \sum_{A=63}^{209} \sigma_\mem{A}
Y(^\mem{A}S) .
\end{equation}
Neutron captures will occur repeatedly on individual
\gdr\ particles, thereby simulating the production of increasingly heavy
\spr\ isotopes.
These neutron captures are responsible for the final \spr\ element
distribution of the species represented by \gdr. They do, however, not
change the number abundance of \gdr.
The ratio $(L/G)=Y(\lei)/Y(\gdr)$
is similar to the customary quantity $n_\mem{cap}$ \citep{clayton:68}
and is a measure of the \spr\ distribution (\abb{fig:LG-ncap}). In fact,
if one defines the artificial sink particle G to be the product of a
n-capture on \fese, then $(L/G)=n_\mem{cap}$. With our choice of G
being the product of a n-capture on \nizw, $n_\mem{cap}$ is slightly  
larger than $(L/G)$ for a given neutron exposure
because $n_\mem{cap}$  takes into account the n-captures on isotopes
from \fese\ to \nizw. These n-captures mainly take place during the initial phase
of the neutron exposure phase for $\tau < 0.1 \mem{mbarn}^{-1}$. 

The neutron cross section of the sink particle
\gdr\ depends on the abundance distribution of the particles it
represents. In \abb{fig:t-Nn-sigma} the Maxwellian-averaged sink cross
section at $8 \mathrm{keV}$ according to \glt{eq:sink-cross} is shown
during calculations of the \spr\ nucleosynthesis with the network including
neutron captures on all isotopes up to Pb, starting with a solar abundance
distribution of
trans-iron elements. The variation of  $\sigma(\gdr)$ reflects the
changing abundance distribution of heavy elements. Initially, species
accumulate at the 
stable neutron magic nuclei on the \spr\ path (\srac, \baac, \blac) and
the averaged cross section decreases. As more neutrons are released
the contribution of species in between increases and finally dominates
the sink cross section. 
The choice of a constant sink neutron cross section for the entire
\spr\ simulation is the largest individual uncertainty when
approximating the \spr\ with artificial sink reactions as
described above. However, it turns out
that the error introduced by using a constant sink cross section is
sufficiently small for our purpose. In
\abb{fig:t-tau-all} we show the neutron exposure of calculations
with the sink treatment with three choices of the sink cross section
as compared with \spr\ calculations performed with a network including
neutron captures on all isotopes up to to Pb for the three 
cases of \abb{fig:t-Nn-sigma}. For low neutron exposures 
the influence of the sink cross section is small, because the
neutron density is dominated by \fese\ and the lighter neutron capture
elements. For the \spr\ in the partial mixing zone of stars of solar 
metallicity the most important range for $\tau$ is between $\sim 0.2
\dots 0.5 \mem{mbarn}^{-1}$. We choose for our simulations
$\sigma(\gdr)=120\mem{mbarn}$ which reproduces the neutron exposure from
the \spr\ calculation up to Pb within $10\%$. 

We also introduced a light neutron-sink reaction to take into account
neutron
captures on elements from S to Mn, which are missing in our network. 
For this reaction, which plays only a minor role, we have used the value 
$\sigma_\mem{light}(8\mem{keV}) = 7.36 \mem{mbarn}$ given by
\citet{lugaro:01a}. 

\section{Mixing from hydrodynamic overshoot}
\label{sec:ov}

The hydrodynamical properties of convection inevitably result in some
turbulent mixing into the stable layers adjacent to  convectively unstable
regions. In fact, any
model of convection in the hydrodynamical framework predicts that the
turbulent velocity field decays roughly exponentially inside the stable
layers \citep[e.g.][]{xiong:85,freytag:96,asida:00}. 
A depth- and time-dependent hydrodynamic
overshoot approximation has been used in a number of recent studies
\citep{herwig:97,mowlavi:99,mazzitelli:99,salasnich:99,herwig:99a,cristallo:00a},
with the aim to capture the main consequences of hydrodynamic mixing into
the stable layers induced by convection.

\subsection{Third dredge-up and H/\czw-partial mixing zone}
\label{sec:OV-dup}
It has been shown by \citet{herwig:99a} that overshoot at all convective
boundaries, including the base of the convective envelope and the bottom
of the He-flash convection zone, strongly increases the efficiency of
the third
dredge-up at low core masses. This is required observationally in order
to reproduce the observed C-star luminosity function in the Magellanic
Clouds \citep{richer:81,frogel:90}. Synthetic models of the AGB phase
in which the third dredge-up 
parameter is derived observationally have demonstrated that typically
efficient third dredge-up must take place at core masses as low as
$0.58\msun$ \citep{marigo:96} or at even lower core masses around
0.54\msun\ according to a more recent analyzes by
\citet{marigo:99}. This condition is not met by most stellar
models, including those of 
\citet{mowlavi:99} who considers hydrodynamic overshoot only at the bottom of
the convective envelope (see his
Fig.\,10b for a comparison of third dredge-up efficiencies found by
different
authors). The model grid of \citet{herwig:00c}, which includes
overshoot at all convective boundaries, includes cases (e.g.\ the
2\msun, Z=0.01 case) which cover the low C-star luminosity tail as required
by observations. The third dredge-up properties of AGB models are important
for the \spr\ because third dredge-up is needed to bring the processed
material to the surface. Hydrodynamic overshoot creates a partial
mixing zone at the core-envelope interface with a continuous decrease
of the H/\czw-ratio from the envelope into the intershell layers
\citep{herwig:97,herwig:99a}. 

The global properties of the \spr\ can be reproduced with a partial
mixing zone resulting from hydrodynamic overshoot
\citep{goriely:00,lugaro:00a}. The main features of the \spr\
overabundance distribution are mainly determined by the regions of
the pocket which have the largest neutron exposures, not so much by the
detailed shape of the H/\czw-profile within the partial mixing zone.
Even if the treatment of hydrodynamic overshoot according to an
exponentially decaying velocity field is not correct and the actual
functional dependence of overshoot efficiency with depth is somewhat
different, then the \spr\ will most likely be affected only slightly as
long as the H/\czw-profile is somehow continuous. For example,
\citet{denissenkov:02} investigate gravity waves below the convective
envelope as a cause for extra mixing to produce partial mixing of
protons and \czw. This mixing process is another way of looking at the
mixing resulting from the perturbation of the convective boundary due
to turbulence and leads to neutron exposures in the region close to that
of previous models featuring a continuous decrease of the proton
abundances into the \czw-rich core. 

An additional effect is introduced in models that consider
overshoot at all convective boundaries \citep{herwig:99a}. In these
models the \czw\ abundance in the intershell is about twice as large  
than in models without overshoot at the base of the He-shell flash convection
zone. \citet{lugaro:03a} have shown that the neutron exposure in the
\spr\ layer is proportional to the \czw\
abundance in the intershell. Hence, in models that consider
overshoot at the base of the He-shell flash convection zone the neutron
exposure in the \spr\ layer is higher than in models that do not include this
overshoot.

\subsection{How much overshoot?}

The initial computations of AGB stars with hydrodynamic overshoot
were carried out with an exponential overshoot parameter 
of $f=0.016$, which was
motivated by the efficiency derived from convective core overshoot
of main sequence stars. The effective mass of the
partial mixing zone where the neutrons are efficiently released is confined within
the region where the proton
abundance follows $-2 < \log X_\mem{P} < -3$ \citep[for an intershell
\czw\ mass fraction of $\sim 20\%$, ][]{goriely:00}. According to this
criterion the mass of the \spr\ layer computed with $f=0.016$ is only
$\sim
10^{-6} \msun$, which is much smaller than required (see
\kap{sec:obs-pm}). 

However, one overshoot efficiency parameter will
not  apply to all convective boundaries during all evolutionary phases. After
\citet{shaviv:73} first considered the possibility of convective
overshoot several studies have used a very simple prescription
in which convective mixing was treated instantaneously and overshoot
was simply a matter of extending the instantaneously mixed region by
some fraction of the pressure scale height. In this approximation
main sequence core overshoot should extend by  
about $0.2\mem{H}_\mem{p}$ \citep[e.g.][ and references
there]{schaller:92}.
\citet{alongi:91} argued that overshoot of $0.7\mem{H}_\mem{p}$
below the envelope of red giant stars could align the location of
luminosity bump with observations. The 2D
radiation-hydrodynamic simulations by \citet{freytag:96} have shown
that the shallow surface convection zone of white dwarfs have exponential
overshoot mixing according to an overshoot parameter 
of $f=1.0$
while the convection zone simulation of A-stars show $f=0.25$. For the
oxygen burning layer in 
pre-supernova models \citet{asida:00} found perturbations of the
stable layers reaching one pressure scale height beyond the formal
convective boundary. Thus, there is ample indication that the
overshoot efficiency is not the same at different convective
boundaries. However, convective overshoot 
is not a stochastic process as long as the convective turn-over time scale
is shorter than the thermal time scale of the region which hosts the
convective boundary. For similar convective boundaries one
should expect a similar overshoot efficiency. This expectation
is supported by 2D hydrodynamical computations by \citet{deupree:00}
 who showed that the core overshoot distance of ZAMS stars varies only
mildly with stellar mass.

Here we choose an exponential overshoot parameter 
 for the
hydrodynamic overshoot at the bottom of convective envelope of
$f=0.16$. This larger overshoot is only applied during the
third dredge-up phase. This has no major effect on the 
properties of the models, other than stretching the partial mixing zone and
consequently the \cdr\ and \nvi\ profiles in that layer over a larger
mass range. The peak neutron exposure and the \spr\ abundance
distribution in the partial mixing layer are not much changed.  
As a side effect the third dredge-up efficiency
is slightly increased, by $\sim 20\%$. 

As mentioned in the introduction observations as well as the analysis
of pre-solar meteoritic SiC grains suggest that 
stars with otherwise identical initial conditions have a range of
\spr\ efficiencies. Such a range can
not be expected to result from overshoot since such
mechanism is not expected to be a stochastic process.

\subsection{Neutron production for the \spr\ in the overshoot model}
\label{sec:pp-ov}

We model the abundance evolution in the partial mixing zone during the
7$^\mem{th}$ interpulse phase of the 3\msun, $Z=0.02$ sequence of
\citet{herwig:00c}, with an overshoot efficiency $f=0.16$ during the third
dredge-up phase. We use the post-processing code (SBM6) described
in \kap{sec:models} because the computation of the stellar evolution does not
include all the species and reactions needed to study the \spr. As initial
conditions we use the thermodynamic and abundance profiles from the stellar
evolution model at the end of the third dredge-up phase after the
thermal
pulse. These profiles are mapped to the equidistant, Lagrangian post-processing
grid and then evolved according to the stellar structure models at a series of
times throughout the interpulse phase. 

We start the simulation with the partially mixed H/\czw\
zone of $\sim 10^{-4}\msun$ that has formed at the end of the
third dredge-up phase as a result of time- and depth-dependent
hydrodynamic overshoot (top panel, \abb{fig:OV-prof}). 
In this model no mixing takes place during the interpulse phase. In
the middle panel of \abb{fig:OV-prof} the \cdr\ neutron source has
started releasing neutrons and up to $10\%$ of the \cdr\ abundance
has been consumed. In the upper part 
of the partial mixing zone where \nvi\ dominates the majority of
neutrons is absorbed by the $\nvi(n,p)\cvi$ reaction. This can be
seen from the profile of \cvi. The maximum neutron density is
$5.7 \cdot 10^{6}\mem{cm}^{-3}$ at
that time and it subsequently reaches a peak value of $1.1 \cdot
10^{7}\mem{cm}^{-3}$.
The bottom panel of \abb{fig:OV-prof} shows the total neutron exposure
in the \spr\ layer before it is
engulfed by the subsequent He-flash convection zone. A maximum neutron exposure
$\tau = 0.72\mem{mbarn}^{-1}$ is found. This is about a factor two
larger  than in previous \spr\ simulations of the partial mixing
scenario \citep{goriely:00} because in our overshoot models the \czw\
intershell abundance is larger than in models that do not use overshoot at the
bottom of the He-flash convection zone. 

In comparison to the observational constraints for the partial mixing
and \spr\ layer derived in \kap{sec:obs-pm} we find that the neutron
exposure in this model with hydrodynamic overshoot at all convective
boundaries is too large. \citet{lugaro:03a}
concluded that to match observed stellar \spr\ abundance patterns the
overshoot at the bottom of the He-flash convection zone
should be smaller than that used by \citet[][ f=0.016]{herwig:99a}. 
For example, the higher temperatures in the He-flash convection zone due
to overshoot result in very large \zrse/\zrvi\ ratios, in
contradiction with measurements in mainstream pre-solar meteoritic SiC
grains. The temperature in the He-flash convection zone decreases with the
stellar mass. However, some SiC show almost no presence of \zrse\ and they 
are
difficult to explain by current models even by considering stars with mass as low as 1.5
$\msun$
\citep{lugaro:03b} and no overshoot included. The inclusion of
overshoot at the base of the He-flash convection zone increases the temperature
for a given stellar mass, making it even more difficult to explain these
data. 
On the other hand, current models of hydrogen-deficient post-AGB stars
\citep{herwig:99c} can only reproduce the 
abundance patterns observed in PG1159-type objects and central stars of planetary 
nebulae of spectral type $\mem{[WC]}$ \citep{koesterke:97b,dreizler:96} 
with AGB progenitor models including intershell overshoot. 
Further investigations of extra mixing have to address
whether the constraints from \spr\ branchings and from 
subsequent evolutionary stages can be resolved in well adjusted
models. 
0
The mass of the layer in which significant overproduction factors
of \spr\ elements can be expected is in our calculation of the
order $\sim 3 \cdot 10^{-5}\msun$, about a factor of two less than the
minimum
pocket mass required according to the estimates in \kap{sec:obs-pm}. 
The choice of the overshoot parameter at the base of the convective
envelope only affects the extent in mass of the
partial mixing zone, and very little the \spr\ distribution.
The latter is determined by the neutron exposure in the pocket.
For that reason one can not
remove the problem of the large neutron exposure in our overshoot model by
a fine tuning of the overshoot parameter at the base of the convective
envelope.

This can easily be shown with a numerical test carried out with a parametric
\spr\ code, as that used by \citet[][and ref.\ there]{lugaro:03a}. 
We made four runs of the \spr\ model with radiative \cdr\
burning through 13 thermal pulse events and mixing episodes according to
the stellar evolution model including overshoot, as described in
\citet{lugaro:03a}. To select the effect of varying of the
\cdr-pocket mass only, we neglect the contribution of the \nezw\ neutron source
during the He-shell flash. The benchmark (BM) case corresponds to the
\cdr\ pocket obtained with an exponential overshoot parameter of
$f=0.128$ at the bottom of the convective envelope.
The pocket is computed starting from an intershell \czw\ mass fraction
of $0.43$ and is kept identical in all the interpulse phases. We
multiply/divide the mass of the pocket as indicated in
\tab{tab:pocket-mass-test}, which approximates the effect of
larger/smaller $f$ values. Note that this is different than
changing the \cdr\ abundance in the pocket as done in order to obtain a
spread in neutron exposures by \citet{busso:01a}.
\tab{tab:pocket-mass-test} gives \spr\ indices at the stellar surface
at the end of an AGB model sequence.  In all cases [hs/ls]$>0$,
clearly in contrast to the observed negative [hs/ls]. The smallest 
[hs/ls] values which are closest to the observed values correspond to
almost no \spr\ overproduction ([ls/Fe] and [hs/Fe] $\sim 0$) and must
be excluded.

\section{Mixing in rotating AGB stars}
\label{sec:rotmix}

For the formation of the \cdr\ pocket and the \spr\ in TP-AGB stars the most
relevant rotationally induced mixing instability  is caused by shear at
locations of large angular velocity gradients. In order to anticipate
rotationally induced mixing events it is therefore important to consider
the angular velocity evolution throughout the thermal pulse
cycle. Throughout this and the following section we make use of the
interpulse phase $\phi = \frac{t-t_0}{\Delta t}$, where $t_0$ 
is the time of the thermal pulse, i.e.\ the peak He-burning
luminosity, and $\Delta t$ is the interpulse period of $3\cdot 10^{4}\jahre$.

The evolution of angular velocity in the intershell region during the 25$^{\rm
th}$ interpulse period of our
$3\msun$ rotating TP-AGB model star is shown in
\abb{fig:OME-interpulse-fig}. The main features of the angular
velocity evolution can be understood by redistribution of angular
momentum in the convective regions and 
conservation of angular momentum in  contracting or expanding stable layers.
The first profile at $\phi=-0.0536$ shows a model of stable
H-burning during the interpulse phase shortly before the He-shell
flash. The H-shell is located just 
below the mass coordinate $0.746\msun$. The steep jump at $\sim
0.741\msun$ marks the largest extent of the He-flash convection zone
during the previous thermal pulse. The He-flash convection zone
redistributes angular momentum from the faster rotating C/O-core
throughout the intershell and the layers just below the
H-rich envelope. The profile shows a plateau
between $0.741$ and $0.746\msun$. This region contains the H-shell
ashes which have been deposited there by the outward burning
H-shell. The 
angular velocity is larger in this area than in the convective
envelope above the H-burning shell. Angular momentum is efficiently
distributed by convection throughout the convective envelope, which 
rotates almost rigidly. However, the H-burning ashes contract 
onto the core and therefore the angular velocity increases, hence the plateau
between $0.741$ and $0.746\msun$ is formed.

At $\phi=0.0034$ the He-flash convection zone has formed and the mass
layers covered by convection are spun up. At this time the He-flash
convection zone reaches up to mass coordinate $0.745\msun$. Between
this and the following profile at $\phi=0.0037$ the He-flash convection
zone reaches its fullest extent up to $0.746\msun$ (just below the
location of the H-burning shell before the thermal pulse). In addition,
the intershell region is expanding due to the energy provided by the
He-flash. This leads to a reduction of the angular velocity in
the intershell. This effect is still present at $\phi=0.0087$, 
until contraction resumes and the angular velocity in the intershell
region increases accordingly (profiles $\phi=0.0122$ to $0.1706$). The
angular velocity profiles at later times show again the formation of a
plateau between $0.746$ and $0.749\msun$, where the H-shell is now located, 
due to deposition of H-shell ashes on top of the core. 

From the evolution of the angular velocity in rotating
TP-AGB stars we can anticipate the following mixing
properties. Efficient shear mixing will take place at $\sim
0.746\msun$ after the formation of the large angular velocity
gradient. This gradient forms because two convective regions extend
to this mass coordinates from below and above in short
succession. The He-flash convection first taps the reservoir of
high angular momentum in the core. Then, the convective envelope 
establishes contact between the fast-rotating intershell and the low-rotating 
envelope. At this interface rotationally induced
shear mixing will be most efficient in 
acting on abundance gradients. The timescale to establish a partial
mixing zone of hydrogen and \czw\ as needed for the formation of the
\cdr\ pocket is limited to the time interval when the envelope and the
intershell are in contact, i.e. between the end of the third
dredge-up and the reignition of the H-shell. This period lasts $2000$
to $3000\jahre$ (\abb{fig:schematic}). However, as
can be seen in \abb{fig:OME-interpulse-fig} the steep angular velocity
gradient remains at this mass coordinate after the formation of the partial 
mixing zone and shear
mixing at this location will persist throughout the intershell period.

Correspondingly, two mixing periods during the interpulse
phase resulting from rotation can be distinguished as shown by the 
temporal evolution of the mixing coefficient presented in \abb{fig:Drot}.
During the initial envelope-core contact period at the end  of the third 
dredge-up a partial mixing zone of protons and \czw\ forms (top panel,
Figure \ref{fig:ROT-prof}). The mixing efficiency in Eulerian
coordinates at this time is   $\log
D_\mem{r} \sim 5 $ (in cgs units), but decreasing  rapidly  at a time scale of
$\sim 1000\jahre$.  However, after the thermal pulse the mass gradient
in the partial mixing zone increases steadily as the intershell layers
contract and evolve towards a pre-flash structure. For assessing the
effect on the  abundance 
evolution the Lagrangian mixing coefficient should be considered:
\begin{equation}
D_\mem{m}=\left({\frac{dm}{dr}}\right)^2 \,
D_\mem{r}=(4\pi\,\rho\,r^2)^2\,D_\mem{r}\, ,
\end{equation}
where all symbols have their usual meaning. 

In \tab{tab:dmdr} we give
the mixing properties in the partial mixing zone at three different times: an 
early phase soon after the pulse, a phase just before the release of neutrons 
starts, and a phase after the neutron source \cdr\ is exhausted.
Despite the comparatively large mixing coefficients rotationally induced
mixing is not important immediately after the initial formation of the partial
mixing zone.
By the time the neutron source is about to be activated the mass
gradient has increased by more than two orders of magnitude, which
more than offsets the decrease of the geometrical mixing coefficient.
The Lagrangian mixing range then exceeds the mass of the initial
partial mixing zone. This trend continues throughout the time of occurrence of 
the \spr\ during the interpulse phase, as the intershell layer and the partial 
mixing zone gradually contract. 

Finally, we note that the present model with rotation does not show
sufficient third dredge-up. This has already been reported
by \citet{langer:99} and is evident from their Fig.\,3 and 4.
We found that the third dredge-up in models of rotating AGB stars is related
to the efficiency of $\mu$-gradients to inhibit mixing. This is a
somewhat uncertain parameterized property of rotating stellar models and a
different approach in describing this effect could change the resulting third
dredge-up.

\subsection{Neutron production for the \spr\ with rotation}
\label{sec:m3r250}

We model the abundance evolution in the partial mixing zone during the
25$^{\rm th}$ interpulse phase of a stellar model of $3\msun$, $Z=0.02$, and
initial equatorial rotation velocity of $250 \mathrm{km/s}$ \citep{langer:99}.
The core mass at this interpulse phase is $M_{\mathrm{c}}=0.746\msun$
and the third dredge-up only just dips into the carbon-rich intershell. 
To compute the \spr\ we used the post-processing code (SBM6) described in
\kap{sec:models}. As in the computation with overshoot, we use as
initial conditions 
the thermodynamic and abundance profiles from the stellar
evolution model at the end of the third dredge-up.
We map these profiles to the equidistant, Lagrangian post-processing
grid and then evolve them according to the stellar structure models at a series
of times throughout the interpulse phase.
As a result of shear mixing due to the large angular velocity gradient 
immediately after the end of the third dredge-up a partial
mixing zone forms (top panel, \abb{fig:ROT-prof}) similar to that in models
with diffusive overshoot (\abb{fig:OV-prof}). The partial
mixing zone defined as having a proton abundance of  $-2 <\log X(\p) <
-3 $  has a mass of $M_\mem{PM} = 6 \cdot 10^{-6}\msun$. 

As described above, throughout the interpulse period shear mixing at the
former core-envelope interface continues at a low level. This has two effects
on the \cdr\ initially formed in the partial mixing zone. The \cdr\ is
spread out, hence diluted, over the mass range subject to shear mixing. 
Secondly, the \nvi\ and the \cdr\ pockets, which are well separated in the
overshoot model, are effectively  mixed. As a result the \nvi\
abundance exceeds the \cdr\ abundance everywhere in the pocket 
(middle panel, \abb{fig:ROT-prof}). In this situation most
neutrons released by \cdr($\alpha$,n)  are absorbed by the reaction
$\nvi(\n,\p)\cvi$ rather than by heavy nuclei. 
Due to the increased amount of neutron absorbers the
neutron density is drastically reduced; the maximum neutron density
is $N_\mem{n}=2.6 \cdot 10^{6}\mem{cm}^{-3}$.
This corresponds to the very small neutron exposure obtained in this
model (bottom panel, \abb{fig:ROT-prof}). 
The activation of the $\nvi(\n,\p)\cvi$ reaction can be seen from the $\cvi$
profile in the middle panel. Protons released by this 
reaction are captured by \czw, building more \cdr, as well as by \cdr\ itself 
(and by \oac, see discussion in \kap{sec:f19}).

In comparison to the basic requirements for the \spr\ layer derived
in \kap{sec:obs-pm} we conclude that this particular model of
rotational mixing in the TP-AGB interpulse period can not produce the
\spr\ abundance patterns observed in stars of solar metallicity. The
neutron exposure reaches only $\tau=0.04\mem{mbarn^{-1}}$, too small for
any significant \spr\ enhancement. As far
as the mass of the \spr\ layer is concerned we note that shear mixing
has caused a substantial broadening. In the end, a zone of $5 \cdot
10^{-5}\msun$ has experienced a low neutron irradiation.

One peculiarity of \spr\ simulations including mixing throughout the
interpulse period concerns the sensitivity to uncertainties of the
nuclear reaction rates. In the parameterized \spr\ without mixing
during the interpulse each layer is treated individually and all \cdr\
is burnt before the onset of the next He-shell flash. As a result, for
example a different $\cdr(\alpha,\n)\ose$ rate, as long as it allows 
all \cdr\ to be consumed during the interpulse period, would lead to a
slight time shift of the neutron release, but in the end this does not
affect the neutron exposure. The
case with mixing throughout the interpulse is different due to
the more important recycling of neutrons by $\nvi(\n,\p)\cvi$ and 
$\czw(\p,\gamma)\ndr$. For
example, we find that the neutron exposure increases by about $20\%$
when using the NACRE \citep{angulo:99} $\cdr(\alpha,\n)\ose$ rate instead of
the CF88 \citep{caughlan:88} rate.

In summary, we have demonstrated the main properties of partial mixing layers
for the \spr\ due to rotationally induced mixing and due to hydrodynamic
overshoot. Both models do not seem to be consistent with the
observed properties. The overshoot model shows
some problems related to the overshoot at the base of the He-flash convection.
The model including rotational mixing shows that the \spr\ during
the interpulse is easily prevented if the delicate process of \cdr\ formation and
neutron release is disturbed by mixing. In fact, our preliminary model of an
AGB star evolving from a main sequence star rotating more slowly than the model 
presented above  
indicates that also in this case the \spr\ nucleosynthesis during the
interpulse period is very much inhibited by the rotation induced mixing. 

\section{Synthetic Mixing Models}
\label{sec:artmix}

In the previous section we have encountered the limitations of both mixing
processes, hydrodynamic overshoot and rotationally induced  mixing. Here
we want to explore the possibility that
some combination of both processes may provide a satisfactory mixing law
for the 
\spr. To that end we construct some synthetic post-processing models
that include overshoot and rotation in a simple parametric
scheme. 

We approximate the details of rotationally induced mixing by assuming a
constant mixing coefficient $D_\mem{IP}$ throughout the
interpulse phase for the post-processing simulation of the \spr\
layer. To show that this approach 
leads to the same results as the detailed computation presented in
\kap{sec:m3r250} we compare the evolution of the neutron exposure with time
from the detailed computation with that obtained in two test cases
(\abb{fig:taupeak}). Test \texttt{rot1} is computed by starting from the
partial mixing zone obtained in the rotation sequence (top panel,
\abb{fig:ROT-prof}), but excluding any further mixing during the interpulse.
The maximum neutron exposure is $\tau \approx
0.38\mem{mbarn}^{-1}$, much higher than the case when mixing continues
during the interpulse and very similar to that found in other partial mixing
models without rotation. However, the neutron exposure integrated over the
simulated mass range is two times smaller in test case \texttt{rot1} compared
to the case when the further mixing during the interpulse is included. 
This reflects the higher degree of n-recycling in a region of wider mass in
the case when rotation mixing during the interpulse is included. Test 
\texttt{rot2} 
is computed by starting again from the
partial mixing zone obtained in the rotation sequence and applying a constant
Eulerian mixing coefficient $\log D_\mem{IP} = 1$ (cgs units) across the
simulation range. This procedure 
can approximately reproduce the key quantities of the detailed post-processing
simulation presented in \kap{sec:m3r250}: $M_\mem{MP} \leq
10^{-4}\msun$ and $\tau_\mem{max} \sim 0.03 \mem{mbarn^{-1}}$.  The fact that
the test case \texttt{rot2} approximates the results of the detailed
simulation confirms that a constant mixing coefficient throughout the
interpulse period mimics the second phase of rotationally
induced mixing that is responsible for the admixture of the \nvi\ neutron
poison and the dilution of the \cdr\ pocket.

With this simple representation we study the effect that 
mixing of the \spr\ layers  during the interpulse might have on the
partial mixing zone formed as a result of hydrodynamic overshoot. We
perform another set of simulations in which we 
start with the partial mixing layer produced from hydrodynamic overshoot
(shown in the top panel of \abb{fig:OV-prof}). A constant mixing coefficient
is imposed during the interpulse simulation. Cases for four different
interpulse mixing efficiencies are shown in \abb{fig:OVROT-tau}. Even with
very weak constant mixing of $\log D_\mem{IP}= -2$ the peak neutron exposure
is somewhat reduced compared to the overshoot case without interpulse mixing
shown in \abb{fig:OV-prof}. For faster interpulse mixing the \spr\ layer
becomes broader and the final neutron exposure declines.

From these tests we tentatively propose a two-step-mixing scheme for
the \spr\ in AGB stars. At the end of the third dredge-up a fast
mixing process induced by the convectively unstable envelope leads to
the formation of a partial mixing zone. A 
mass range of $\approx 5 \cdot 10^{-5}\msun$ should have a proton abundance
of $-2 < \log X(\p) < -3 $. The intershell \czw\ abundance should be larger
than in models that confine mixing in the He-shell flash convection
zone strictly  to within the Schwarzschild boundaries. Due to
mixing during the interpulse phase the neutron release is
modulated (in any 
case reduced) and a set of otherwise identical stars with a variety of mixing
efficiencies at the core-envelope layers will display a spectrum of neutron
exposures. In this scenario involving hydrodynamic overshoot as well
as interpulse mixing, 
the efficiencies needed to cover the spread in \spr\ efficiencies observed in
stars \citep{busso:01a} and pre-solar grains \citep{lugaro:03b} can be achieved
with mixing efficiencies of the order $\log D_\mem{IP} \approx 0 \dots -1.3$.  
This range is not currently reproduced by stellar evolutionary
sequences including rotation, which predict much larger mixing
efficiencies (see \kap{sec:m3r250}). Smaller 
mixing efficiencies would result from a smaller angular velocity
gradient at the core-envelope interface and/or a smaller rotation
rate. This could be achieved by a more efficient angular 
momentum transport during the progenitor evolution.
 Efficient third dredge-up and/or penetration of  
the convective pulse into the core might affect the mixing properties due
to rotation as angular momentum could be transported out of the core into the
envelope and carried away by mass loss. This effect is not present in our computation as our models do not
show efficient third dredge-up.
Moreover, in our models we have not considered magnetic fields
which could enhance the coupling of core and envelope and decelerate the core
\citep{spruit:98}.

\section{Fluorine Production in  Models with Rotation}
\label{sec:f19}

\citet{jorissen:92} have observed high enhancements of fluorine in
AGB stars. The solar abundance is  $X(\fne)=4.1 \cdot 10^{-7}$    
and observationally $X(\fne)=6.6 \cdot 10^{-7}$  at the start of the 
AGB phase (average of K and M giants). The typical abundance of \fne\
in TP-AGB stars (S, M and C stars) is $X(\fne) \approx (4.5 \pm{4})
\cdot 10^{-6}$. In particular the N-type carbon stars show a large
spread in \fne\ abundances within a small range of C/O values. Overall
thermal pulses appear to cause a tenfold increase in \fne\ as a result of thermal
pulse and/or interpulse nucleosynthesis and third dredge-up. While the
observed correlation of fluorine enhancement with s-process enhancement
is reproduced  by current models \citep[see Figure 13 of][]{goriely:00}, the  
observed correlation of fluorine with carbon remains unsolved.

Several nucleosynthesis paths have been considered and
the most likely involves a (n,p) reaction \citep{jorissen:92}. 
Because the $\fne(\alpha,\p)\nezw$ rate is about ten times
larger than the $\nezw(\alpha,\n)\mgfu$ rate at $T=3 \cdot 10^{8}\kelv$
\citep{caughlan:88,angulo:99} $\nezw$ can be excluded 
as the $n$-source for the required (n,p)-reactions. Even if fluorine
is made during the interpulse period, massive AGB stars can be excluded as
\fne\ producers because they have very hot He-flash convective zones, in which 
$\fne(\alpha,\p)\nezw$ is  very efficient.
Moreover, in massive AGB stars, proton captures at the hot base of the
convective envelope destroy fluorine efficiently.

The \cdr($\alpha$,n) reaction is activated when the H-burning ashes are engulfed
in the convection zone. However, the amount of neutrons released by the \cdr\ 
from the H-burning ashes is
not enough to produce the required abundance of fluorine. Another possible
site for the production of fluorine is the nucleosynthesis occurring in
the partial mixing zone during the interpulse periods.
However, the partial mixing zone is typically unimportant with regards
to the production of light elements because its mass is very small.
In fact, the inclusion of the partial mixing zone in model calculations does
not appear to increase the predicted surface abundance of \fne. This
conclusion can be drawn when comparing the abundance of \fne\ at a given C/O
ratio in Figure 12 of \citet{goriely:00},
who studied fluorine production in AGB stars including a parametric partial
mixing zone, and in Figure 14 of \citet{mowlavi:96}, who did not include a
partial mixing zone in their study. Here we check whether rotation can improve
the match with observation.

During the interpulse period and in the partial mixing zone about
$80\%$ of the \nfu\ is produced via the 
chain of reactions \begin{displaymath}
\nvi(\n,\p)\cvi(\alpha,\gamma)\oac(\p,\alpha)\nfu 
\end{displaymath}
and 20\% comes from the reaction $\nvi(\n,\gamma)\nfu$ (\abb{fig:N15}). 
During the interpulse phase the temperature is not high enough for the 
$\nfu(\alpha,\gamma)\fne$ reaction and the production of
fluorine occurs later in the He-flash convection zone
(note that the $\nfu(\alpha,\gamma)\fne$ rate is about 50 times faster than the
destruction of \fne\ by $\alpha$-capture in the temperature range 
of interest). For simplicity we assume that all \nfu\
present in the intershell before the onset of the He-flash will be
transformed into \fne\ and no \fne\ will be destroyed so that the
intershell abundance at the time of the third dredge-up is
$X_{IS}(\fne) \propto M_{^{15}N} / M_\mem{IS}$, where 
$M_{^{15}N}=X_{PM}(\nfu) M_\mem{PM}$. With 
\glt{math:mix}, the observed abundances mentioned above, using q=2.3,
m=20 and the other numbers used in \kap{sec:obs-pm}, the total \nfu\
production in the partial mixing zone during 
the  interpulse phase must be of the order $M_{^{15}N}> 
10^{-7}\msun$ to cover the observational data.

 In \abb{fig:N15} we show some results for 
the \nfu\ production in the partial mixing zone.
In the post-processing model of the sequence
including rotation \nfu\ is produced by the reaction channels described
above. The absolute amount is much smaller than
needed. However, compared to a case with the same initial partial
mixing zone but no interpulse mixing (test case rot1,
\kap{sec:artmix}) the production of \nfu\ is larger by a factor
$2\dots3$. We analyzed the \nfu\ production in the parametric models 
presented in \kap{sec:artmix} that combine an overshooting partial mixing
zone with interpulse mixing of the \spr\ layer. We find that the \nfu\
production increases with  interpulse mixing reaching $8\cdot 10^{-8}
\msun$  for $D_\mem{IP}=0$.  If rotation in AGB stars is 
instrumental in producing \fne\ one might expect an
anti-correlation of \fne\ with the s-process index [hs/ls] as well as with
the s-process enhancement. Currently available observational data do
not show this.
Renewed observational work on \fne\ in AGB stars is needed as well as more
detailed models. In addition the question of reaction rate uncertainties has to
be revised, in particular in the case of $\cvi(\alpha,\gamma)\oac$.

\section{Discussion}
\label{sec:disc}

We have compared two mixing processes for the \spr\ in low-mass AGB
stars. Models that include overshoot at all convective boundaries
feature neutron exposures that are larger than allowed to reproduce
the observed \spr\ properties. This is due to 
the fact that overshoot at the base of the He-shell flash
convection zone leads to intershell dredge-up of additional carbon
from the core into the intershell layer. 
Models that do not include overshoot at all convective boundaries (and
include the partial mixing zone in a parametric way) feature neutron exposures
that can explain only a small fraction of the observations.
To explain the whole range of observational data some spread in the 
efficiency of the neutron release is required in the scenario of
$^{13}$C($\alpha$,n) occurring in the interpulse periods. If rotation is
included the \spr\ layer located at the core-envelope interface is continuously
mixed throughout the interpulse period. The \spr\ layer is not
stratified anymore and the \cdr\ pocket which forms at the end of the
third dredge-up is polluted with the \nvi\ from layers in the partial
mixing zone with an initially larger H/\czw\ ratio. Using the mixing law
from a stellar evolution sequence of an initially rapidly
rotating star the resulting neutron exposure is too small to allow any
production of \spr\ elements.

We have constructed synthetic models with a parameterized range of mixing
efficiencies during the interpulse phase. In these models we find a
spread of neutron  
exposures in the \spr\ layer. Hence, we tentatively propose that the
signature of rotationally induced mixing in AGB stars might be identified with
the observationally-inferred spread of neutron exposure in the interpulse.
The spread of mixing efficiencies may be related to a spread of angular
velocity gradients in stars of different initial rotational velocities, masses,
or evolutionary status. However, mixing efficiencies which would correspond to
the observed spread of neutron exposures are substantially lower than the
mixing law in our stellar evolution model with rotation. In this scenario this might be
attributed to some effect of angular momentum redistribution and loss
due to a physical process missing in our calculations, such as efficient third
dredge-up, mass loss and magnetic fields.

In the framework of the radiative interpulse \spr\ production
site our simulations establish a relation between the intershell carbon
abundance and the range of mixing efficiencies during the
interpulse phase of AGB stars. Any mixing in the \spr\ layer during the 
interpulse 
phase and after the initial formation of the partial mixing
zone can only reduce the final neutron exposure. A spectrum of slow
mixing efficiencies during the 
interpulse phase can deliver a spread in neutron exposures in
accordance with the observed spread of \spr\ indices if the neutron exposure in 
the limiting case without mixing is larger
than the largest 
observationally required neutron exposure \citep[for example case
$\mem{ST}\cdot 2$ in][]{busso:01a}.  If the 
carbon intershell abundance
is larger than that obtained in models without hydrodynamic
overshooting at the 
bottom of the He-flash convection zone, then the \cdr\ abundance in
the PM zone and
thus the neutron exposure will be larger as well.  As discussed in
\kap{sec:pp-ov} AGB models including this effect are in very good
agreement with the observed carbon abundances of H-free and hot central
stars of planetary nebulae, in which the progenitor AGB intershell
abundance is believed to be seen at the surface.

In this paper we have used the nucleosynthesis and the observed
features of the \spr\ to constrain  stellar
models. We have demonstrated how different mixing processes in
different locations and at different times impact the current model of \spr\
nucleosynthesis in low-mass AGB stars. In the future we should improve the
stellar physics in order to investigate which processes determine the
 mixing at the core-envelope interface and why mixing efficiencies in our
rotating stellar model are not in agreement with the
requirements of the \spr. The theory of angular momentum transport and loss may
need to be revisited. The question of the third dredge-up in rotating AGB
stellar models needs additional consideration, and it is likely that the
treatment of $\mu$ gradients and their impact on mixing will play an
important role. 

\acknowledgments
F.\,H.\ appreciates  support from  D.\, A.\ VandenBerg through his
Operating Grant from the Natural Science and Engineering Research
Council of Canada. F.\,H.\ also thanks the Institute of Nuclear Theory
at the University of Washington for its hospitality and the U.S.\ Department
of Energy for partial support during the completion of this work. 
The detailed and thoughtful comments by the referee have helped us 
to improve the  presentation of this paper.
 
\newpage
\appendix

\section{Details of the computation of [hs/ls] and s/s$_{\odot}$ vs
$\tau$}
\label{app:n14}
In \kap{sec:obs-pm} we have presented a \spr\ calculation
(\abb{fig:tau-shsls})  to establish the relation between the neutron
exposure and the \spr\ indices [hs/ls] and $\log(s/s_0)$.  We have
used for this calculation initial mass fractions for the light neutron
poison typically encountered in the partial mixing zone, including
$(\hevi,\czw,\ose,\nezw) \simeq (0.74,0.23,0.01,0.02)$. In our
calculation the temperature and density are chosen constant at  
$T=9.8\cdot 10^{7}\kelv$ and $\rho=2000\mem{g/cm^3}$.
For intermediate- and heavy-mass isotopes we have used an initial solar
abundance distribution. The charged-particle reactions listed in 
\kap{sec:models} and neutron-capture reactions on isotopes up to Pb are
considered in the calculation. We have taken an initial mass fraction
of $^{13}$C of 0.03 and assumed only a residual amount of initial nitrogen 
($X(\nvi)=1.2\cdot10^{-5}$). The
initial amount of protons was set to zero. With this choice a final
neutron exposure of $\tau \simeq 0.7$ mbarn$^{-1}$ is reached when all 
$^{13}$C is consumed.

In a partial mixing situation a
significant amount of \nvi\ may be present. Due to different
assumptions about mixing in the underlying stellar evolution model,
the abundance of \czw\ and \ose\ may be 
different from model to model. The larger
intershell \czw\ abundance in models with hydrodynamic overshooting
increases the neutron exposure that can be generated in the partial
mixing zone \citep{lugaro:03a}.
 However, we show here that the
relation between $\tau$ and the \spr\ indices does not depend on the
abundance distribution of light poison, for example the presence of
\nvi. We demonstrate this by some test calculations and by recalling
some basic properties of the \spr\ as layed out in text books like
\citet{clayton:68}. 

The neutron exposure is defined as $\tau = \int\!_0^t N_\mem{n}
v_\mem{T} \, dt$, where $N_\mem{n}$ is the neutron density and
$v_\mem{T}$ is the thermal velocity. The neutron density is related to
the molar neutron abundance by $N_\mem{n}=\mem{N}_\mem{A} \, \rho \, 
Y(\mem{n})\, A(\mem{n})$ where all symbols have their usual meaning. 
Since neutron captures are faster
than any of the charged particle reactions the neutron density is
given by the ratio of neutron production by sources and destruction by
sinks. The presence of 
light neutron poison (sinks) will depress the neutron density (and
eventually the neutron exposure) and will thereby determine the amount
of neutrons available to enter the \spr\ path.
The \spr\ distribution (as given by the [hs/ls] index) as well as the
total production of the \spr\ elements (as given by the index
$\log(s/s_0)$) is then given by the number of neutrons that
are captured by the \spr\ seed (usually \fese) and other heavy elements. This
number depends solely on the neutron density and  the abundances and
reaction rates of seed nuclei and all \spr\ species themselves. While
neutron poison may limit the neutron density (and 
eventually $\tau$) they can not directly influence the relation of the
integrated neutron density ($\tau$) and the \spr\ distribution and
total enhancement.

More specifically, the \spr\ distribution depends univocally on the number
of neutrons, $\mem{n_{56}}$, captured by $^{56}$Fe and its progeny. 
If the abundance of
$^{56}$Fe is much 
larger than the abundance of its progeny 
$\mem{n_{56}}$ is equal to the amount of $^{56}$Fe destroyed by neutron
captures:
\begin{equation} \label{eq:n56}
Y(\mem{n_{56}})=-Y(^{56}{\rm Fe})=\int Y(^{56}{\rm Fe})\, Y(\mem{n})\,
r_{56,(n\gamma)}\, dt,
\end{equation}
where $r_{56,(n\gamma)}$ is the rate of the $^{56}$Fe($n,\gamma$) reaction
and $Y$ are molar abundances. 
 If we assume $^{13}$C($\alpha,n)$ to be the neutron
source and $^{56}$Fe($n,\gamma)$ and $^{14}$N($n,p)$ to be the major neutron
absorbers, then with the equilibrium condition $dN_\mem{n}/dt=0$ the
total neutron abundance is given by: 
$$
Y(\mem{n})=\frac{Y(^{13}{\rm C})\, Y(^{4}{\rm He})\,
r_{13,(\alpha,n)}}{Y(^{14}{\rm
N})\, r_{14,(n,p)} + Y(^{56}{\rm Fe})\, r_{56,(n,\gamma)}},
$$
where $r_{14,(n,p)}$ is the rate of the $^{14}$N($n,p)$ reaction and
$r_{13,(\alpha,n)}$ is the rate of the $^{13}$C($\alpha,n)$ reaction.
The abundance of light neutron poison like $^{14}$N determines the
neutron density on which the number of neutron captures by \fese\ and progeny
depends \glp{eq:n56}. But the light neutron poison do not affect the relationship between 
$Y(\mem{n_{56}})$ and $Y(\mem{n})$, i.e.\ between the
\spr\ distribution and the neutron exposure.

 In \abb{fig:tau-N14} we show
different cases of the same computation of \abb{fig:tau-shsls} performed with a 
range of  initial  $^{14}$N mass fractions.
Changing the initial mass fraction of $^{14}$N leaves almost completely
unchanged the relationship between the \spr\ distribution and the neutron
exposure. The difference is in the final $\tau$ value, which is much lower for
a higher $^{14}$N abundance. In the same way the relation of $\tau$ with
the \spr\ indices is independent on the abundance of other light
neutron sinks, and rather general. 

It follows that the parameter $\tau$ determines
univocally the \spr\ distribution. Note, that the  conclusions 
reached in \kap{sec:artmix} do
not change if instead of $\tau$ we quantify the \spr\  with the
parameter $(L/G)$, which was introduced in \kap{sec:sink}, and which 
is a measure of
the number of neutrons captured by the heavy sink particle \gdr, i.e. is
similar to considering the number of neutrons captured by $^{56}$Fe. In the
synthetic models presented in \kap{sec:artmix} for a range of $\log D_\mem{IP}
\approx 0 \dots -2.$ we obtain a  range of $\tau$ and correspondingly
a range of $(L/G)$ (see \tab{tab:taulg}).

\clearpage
\begin{deluxetable}{lllll}
\tablecolumns{5}
\tablewidth{0pc}
\tablecaption{\label{tab:pocket-mass-test}
Effect of the \cdr-pocket mass on the \spr\ efficiency and
abundance distribution}
\tablehead{
\colhead{IP}    &  \colhead{$\mem{BM}/10$}&  \colhead{$\mem{BM}/2$}&
\colhead{$\mem{BM}^{(a)}$}&  \colhead{$\mem{BM}\cdot 2$}
}
\startdata
$\mem{[ls/Fe]}$ &$         0.03$&$   0.13 $&$ 0.23$&$  0.38$\\
$\mem{[hs/Fe]}$ &$         0.06$&$   0.25 $&$ 0.40$&$  0.61$\\
$\mem{[hs/ls]}$ &$         0.03$&$   0.12 $&$ 0.17$&$  0.23$\\
\enddata
\\
\flushleft{{$^{(a)}$}BM: benchmark \cdr\ pocket corresponding to case with
hydrodynamic overshoot with $f=0.128$.}
\end{deluxetable}

\begin{deluxetable}{cccccc}
\tablecolumns{6}
\tablewidth{0pc}
\tablecaption{\label{tab:dmdr}
Mixing coefficients and Lagrangian mixing efficiency in the partial
mixing layer of the AGB model including rotation} 
\tablehead{
\colhead{$\phi$} &  \colhead{$dm/dr^{(a)}$} & \colhead{$\log D_\mem{r}^{(b)}$}&
\colhead{$\log D_\mem{m}$} &  \colhead{$\Delta t_\mem{mix}$} &
\colhead{$\Delta m_\mem{mix}$}\\
 &$\msun /\mem{cm}$   &
[$D_\mem{r}$]$=\mem{cm^2}/\mem{s}$& [$D_\mem{m}$]$=\msun^2/\mem{yr}$ &
$\mem{yr}$& \msun }
\startdata
0.034& $5.4 \cdot 10^{-14}$ & $\sim 5.0$&$-14.0$&$1440^\mem{(c)}$ & $3.7
\cdot 10^{-6}$\\
0.232& $9.2 \cdot 10^{-12}$ & $\sim 1.5$&$-13.1$&$8560^\mem{(c)}$&$2.7\cdot10^{-5}$\\
0.557& $3.6 \cdot 10^{-11}$ & $\sim 2.8$&$-10.6$&$2900^\mem{(d)}$&$2.7\cdot10^{-4}$\\
\enddata
\\
\flushleft{$^{(a)}$at $m_\mem{r}=0.745965\msun$; $^{(b)}$max, see
\abb{fig:Drot} for the overall distribution in mass; $\mem{(c)}$time since 
thermal pulse; $\mem{(d)}$nuclear
lifetime of \cdr\ against $\alpha$-capture $\tau_\alpha(\cdr)$. }
\end{deluxetable}

\begin{deluxetable}{ccc}
\tablecolumns{3}
\tablewidth{0pc}
\tablecaption{\label{tab:taulg}
Maximum $\tau$ and $(L/G)$ obtained for different $\log D_\mem{IP}$
}
\tablehead{
\colhead{$\log D_\mem{IP}$} &  \colhead{$\tau$} & \colhead{$(L/G)$}\\
$[D_\mem{r}]=\mem{cm^2}/\mem{s}$ & $\mem{mbarn}^{-1}$ & number ratio 
}
\startdata
-0.0 & 0.12 & 4 \\
-1.0 & 0.33 & 18 \\
-2.0 & 0.64 & 55 \\
no mixing & 0.72 & 70 \\
\enddata
\\
\end{deluxetable}
\clearpage


\clearpage

\begin{figure}
\plotone{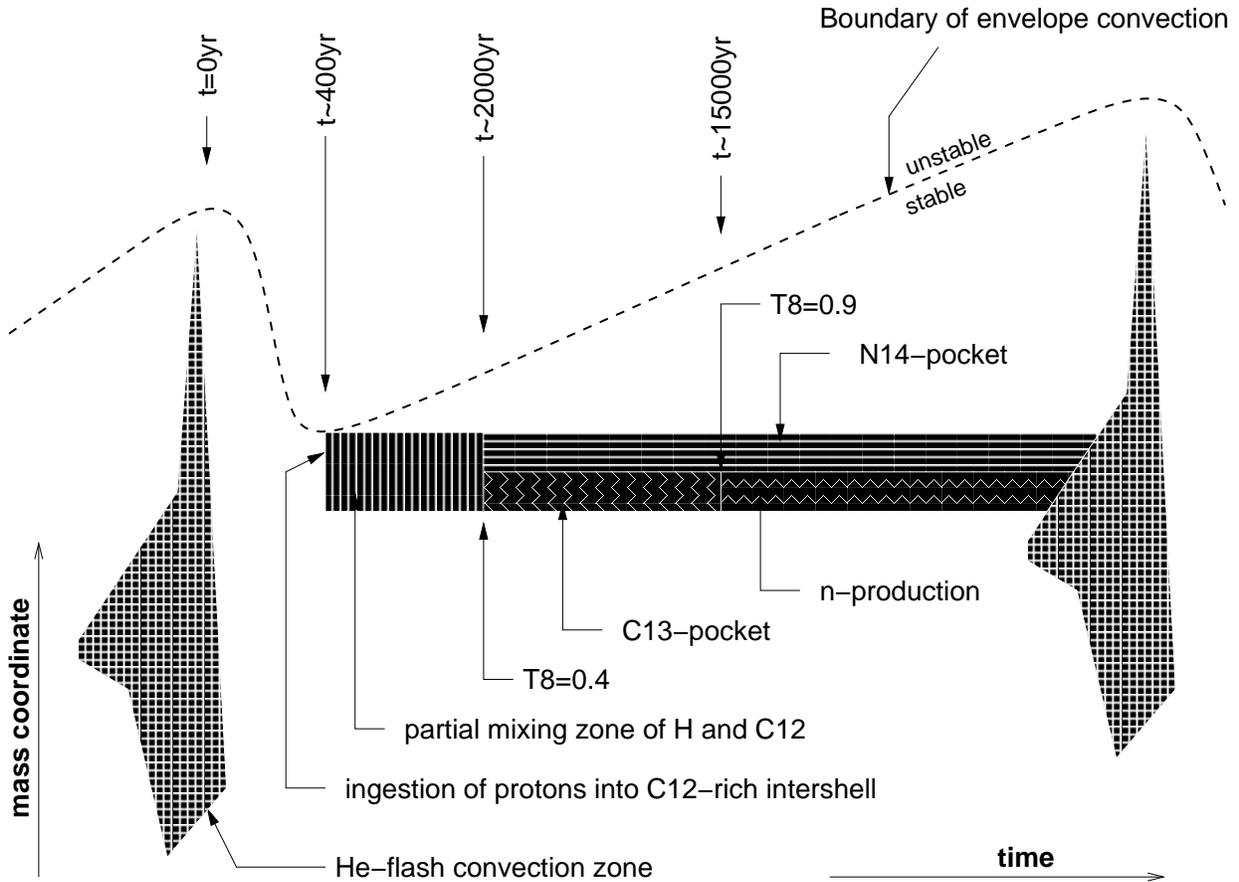} 
\figcaption{ \label{fig:schematic} 
Schematic representation of the \spr\ in the interpulse phase of
TP-AGB stars in a space-time diagram. The ordinate covers the mass
range between the H- and He-burning shells (intershell). The time marks
represent a rough estimate.
}\end{figure}

\begin{figure}
\plotone{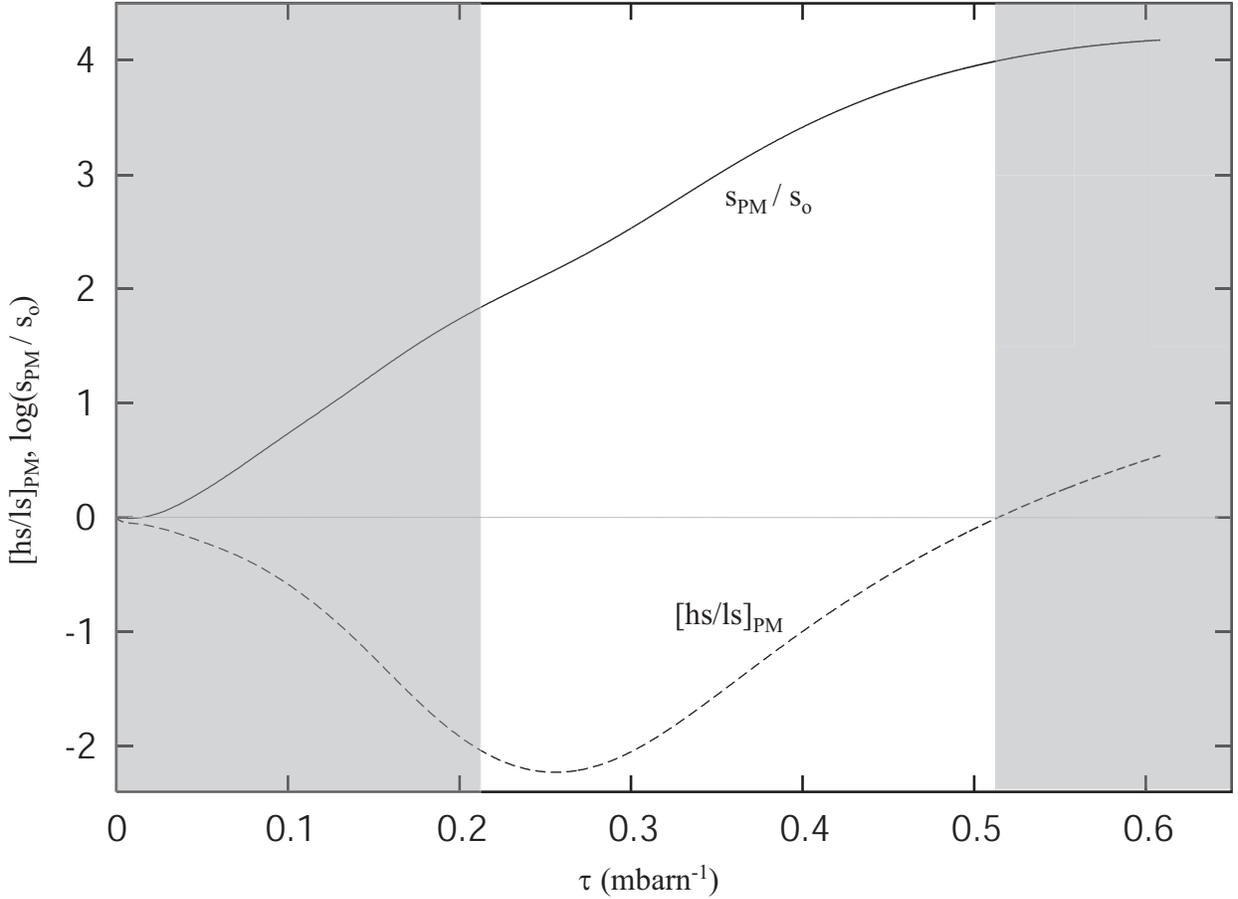} 
\figcaption{ \label{fig:tau-shsls} 
\spr\ indices log($s/s_{\odot}$) and [hs/ls] in  $s$-processed material as
function of the neutron exposure from a network calculation
including all relevant light and heavy elements from H to Pb. 
The initial \cdr\ mass fractions is 0.03 and neutrons are
released by the \cdr($\alpha$,n) reaction at a 
constant temperature of $9.8 \cdot 10^7$ K.  
The neutron exposure corresponding to the shaded areas can be excluded
as the the dominant contribution in the \spr\ site (see
text). As discussed in Appendix A the presence of light neutron poison
like \nvi\  would limit the neutron exposure that can be achieved but
the relations of $\tau$ and the shown indices are only very weakly effected. 
}
\end{figure}
\begin{figure}
\plotone{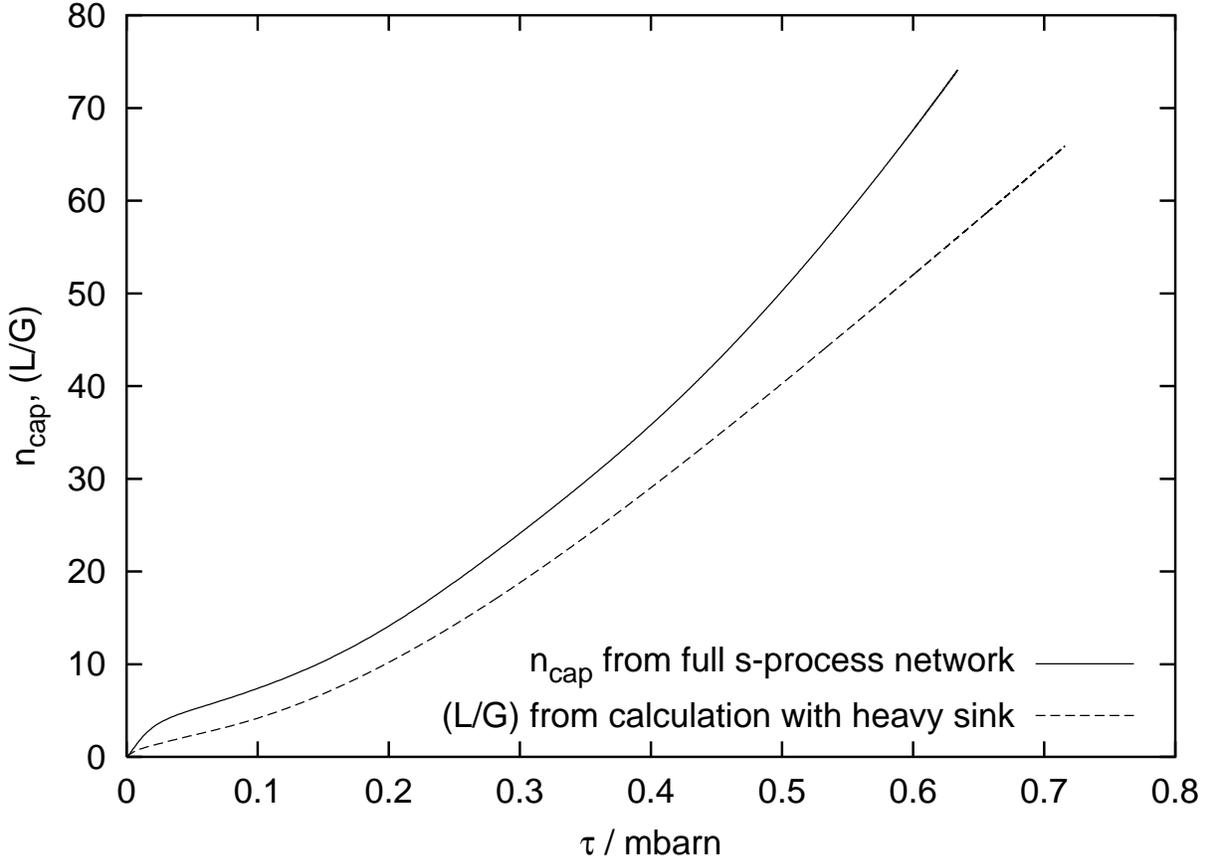} 
\figcaption{ \label{fig:LG-ncap} 
Comparison of the \spr\ parameters $n_\mem{cap}$ and L/G as functions of 
$\tau$. The quantity $n_\mem{cap}$ is computed with a full \spr\
network calculation while the ratio L/G comes from the calculation in which
the \spr\ is approximated with a sink. The first increase of $n_\mem{cap}$ 
with respect to L/G at very low $\tau$ is due to neutron captures on isotopes 
from $^{56}$Fe to $^{62}$Ni. From $\tau \sim 0.5$ mbarn$^{-1}$ the resul;ts 
obtained with the neutron sink show larger and larger differences from the 
results obtained with the full \spr\ network. The neutron sink representation 
is a valid approximation for $\tau < 0.6$ mbarn$^{-1}$, i.e. in the range of 
interest for solar metallicity stars.} \end{figure}

\begin{figure}
\plotone{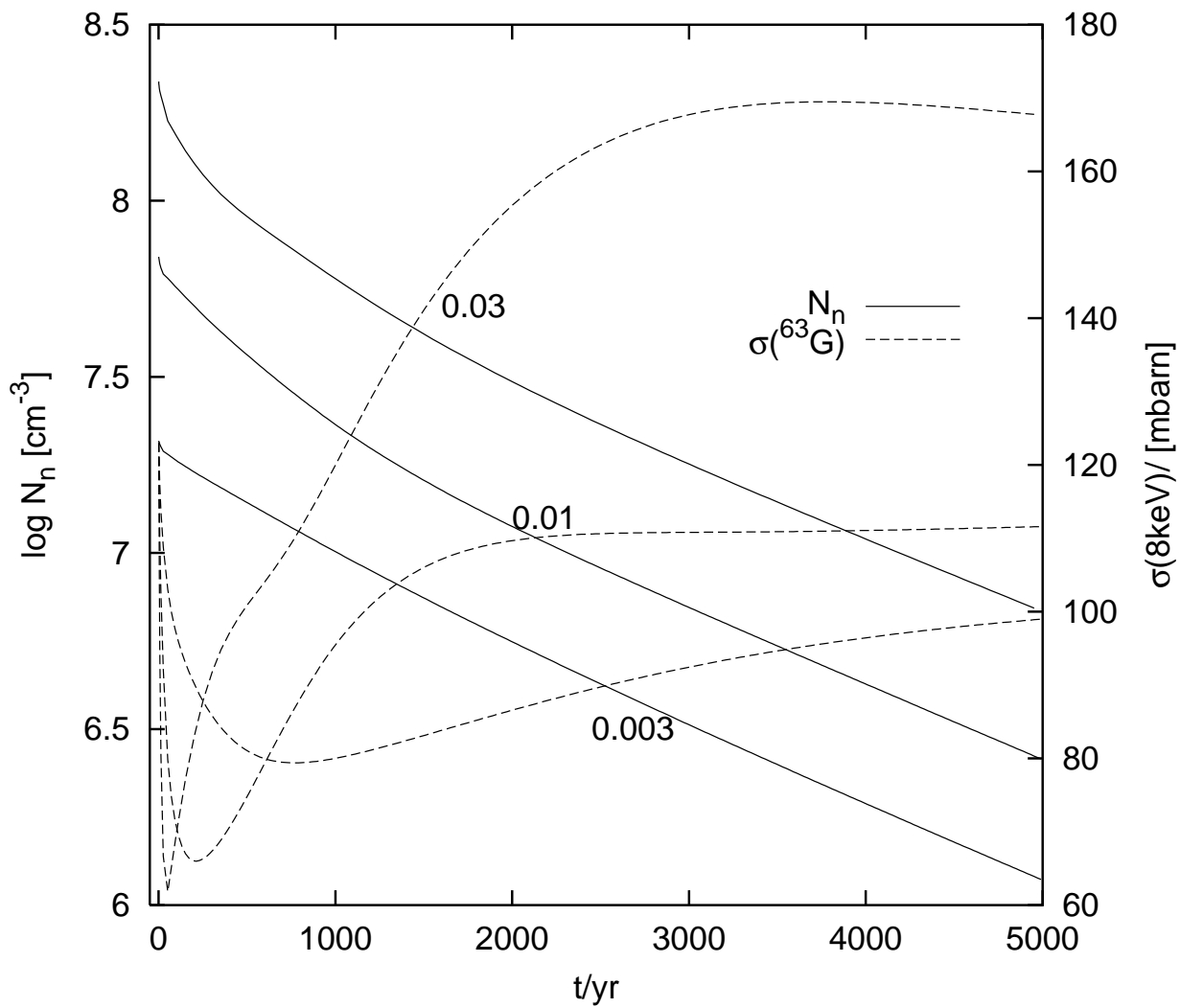} 
\figcaption{ \label{fig:t-Nn-sigma} 
Neutron density (solid line, left ordinate) and Maxwellian-averaged neutron
capture cross section (dashed line, right ordinate) of sink particle
\gdr\ (described in \kap{sec:sink}) as a function of time from \spr\
calculations as those performed for \abb{fig:tau-shsls} with a network including
all heavy elements up to Pb.
Three test calculations are presented starting with three different amount of
\cdr\ mass fractions (indicated by the labels).
The variation of the cross section of \gdr\ reflects its dependence on the
abundance
distribution of the species represented by the sink particle.
}
\end{figure}

\begin{figure}
\plotone{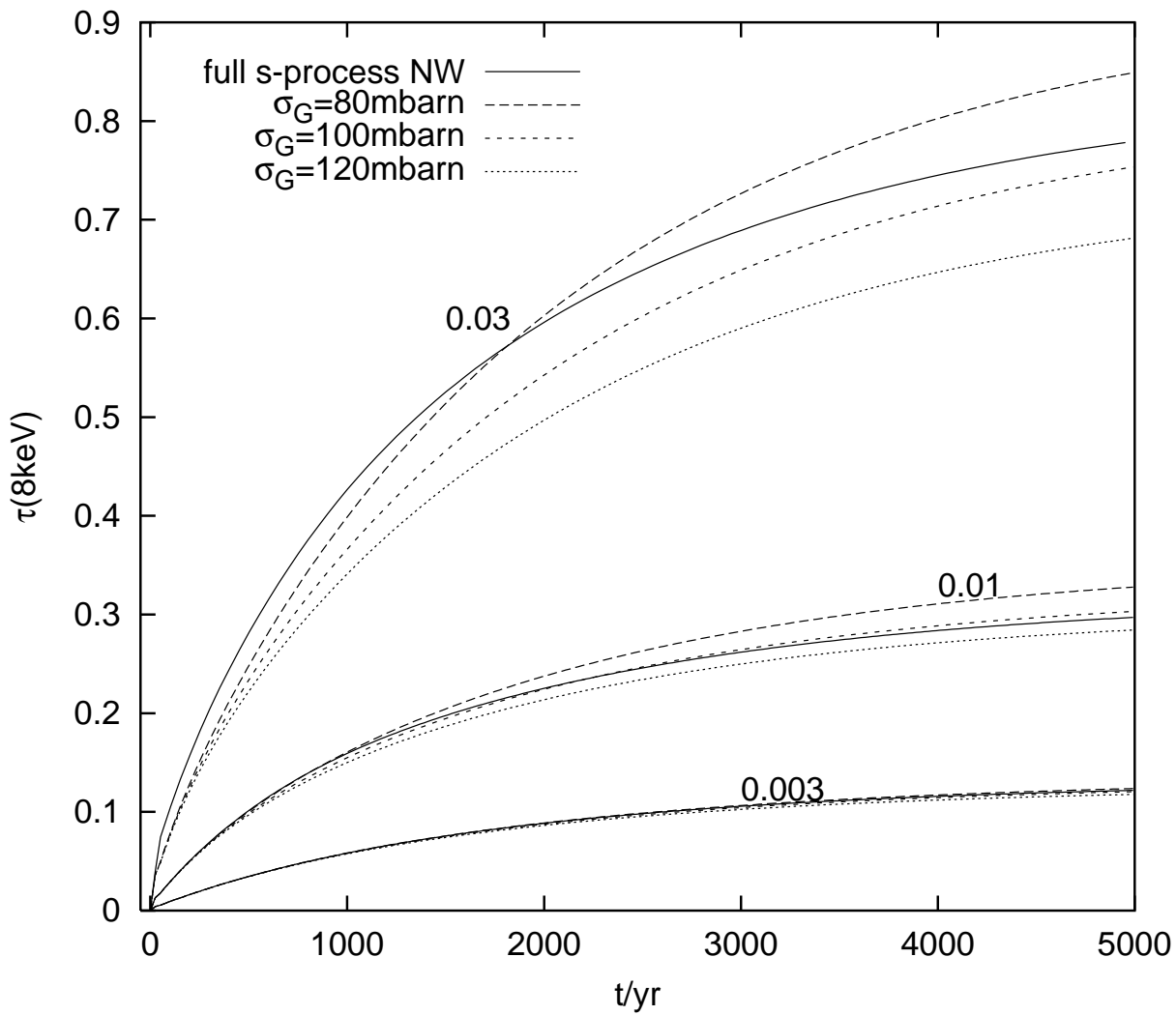} 
\figcaption{ \label{fig:t-tau-all} 
The evolution in time of the neutron exposure $\tau$ for the three calculations
shown in \abb{fig:t-Nn-sigma} is represented by the solid lines (initial
\cdr\ mass fractions
indicated by the labels). Each calculation is compared to three tests
computed with different values of the Maxwellian-averaged neutron capture 
cross section of the neutron sink reaction $\sigma(\gdr)$ described in 
\kap{sec:sink}.
}
\end{figure}

\begin{figure}
\plotone{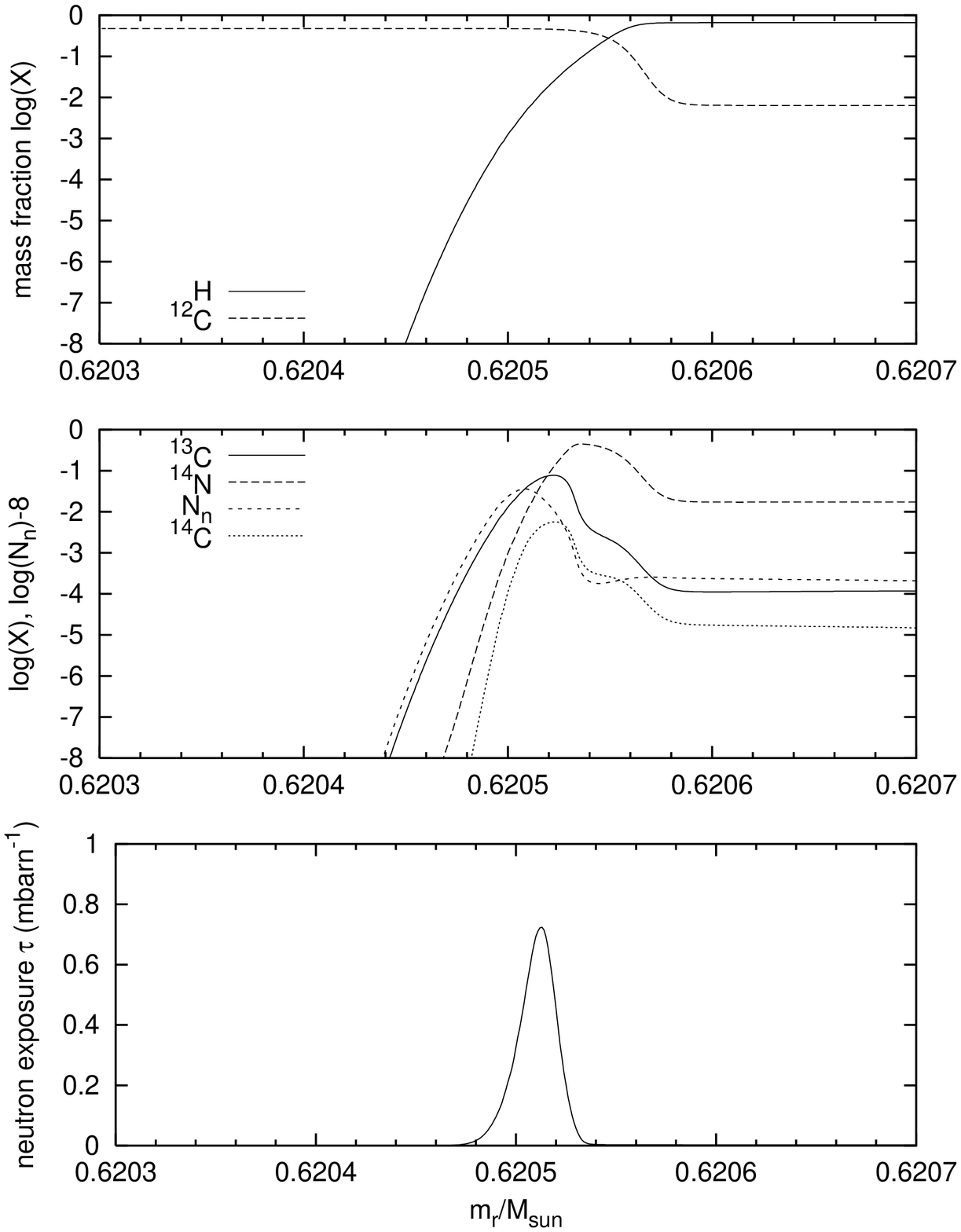}
\figcaption{ \label{fig:OV-prof} for caption see page \pageref{figcap}
}
\end{figure}

\begin{figure}
\plotone{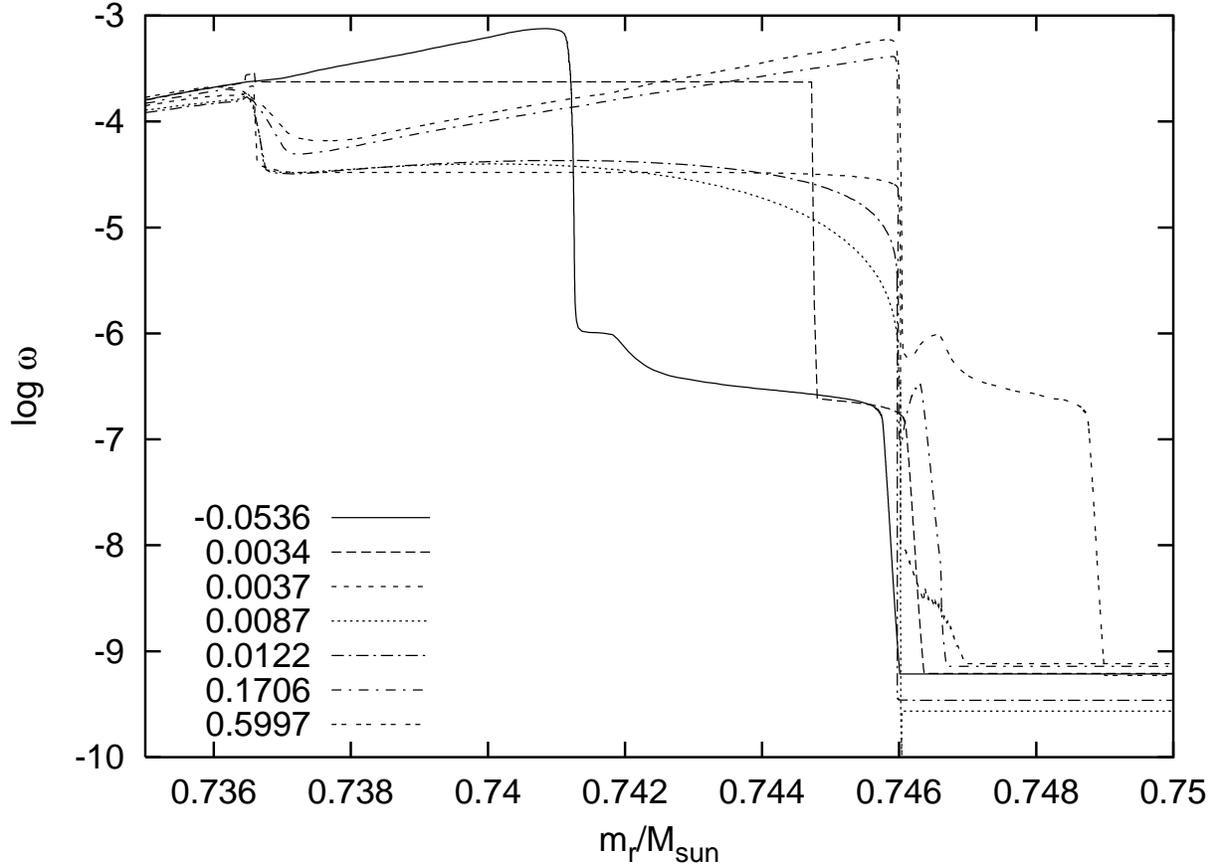} 
\figcaption{ \label{fig:OME-interpulse-fig} 
Evolution of angular velocity profiles through the 25$^{\rm th}$
interpulse-pulse cycle. 
Labels give interpulse phase $\phi=(t-t_0)/\Delta t$, 
where the previous pulse, i.e. the maximum 
He-burning luminosity,
has occured at  $t_0$;  the interpulse period
is $\Delta t=3\cdot 10^{4}\jahre$.  
The mass range covers the intershell region from the
top of the C/O-core (location of the He-shell) to the
bottom of the convective envelope.}
\end{figure}

\begin{figure}
\plotone{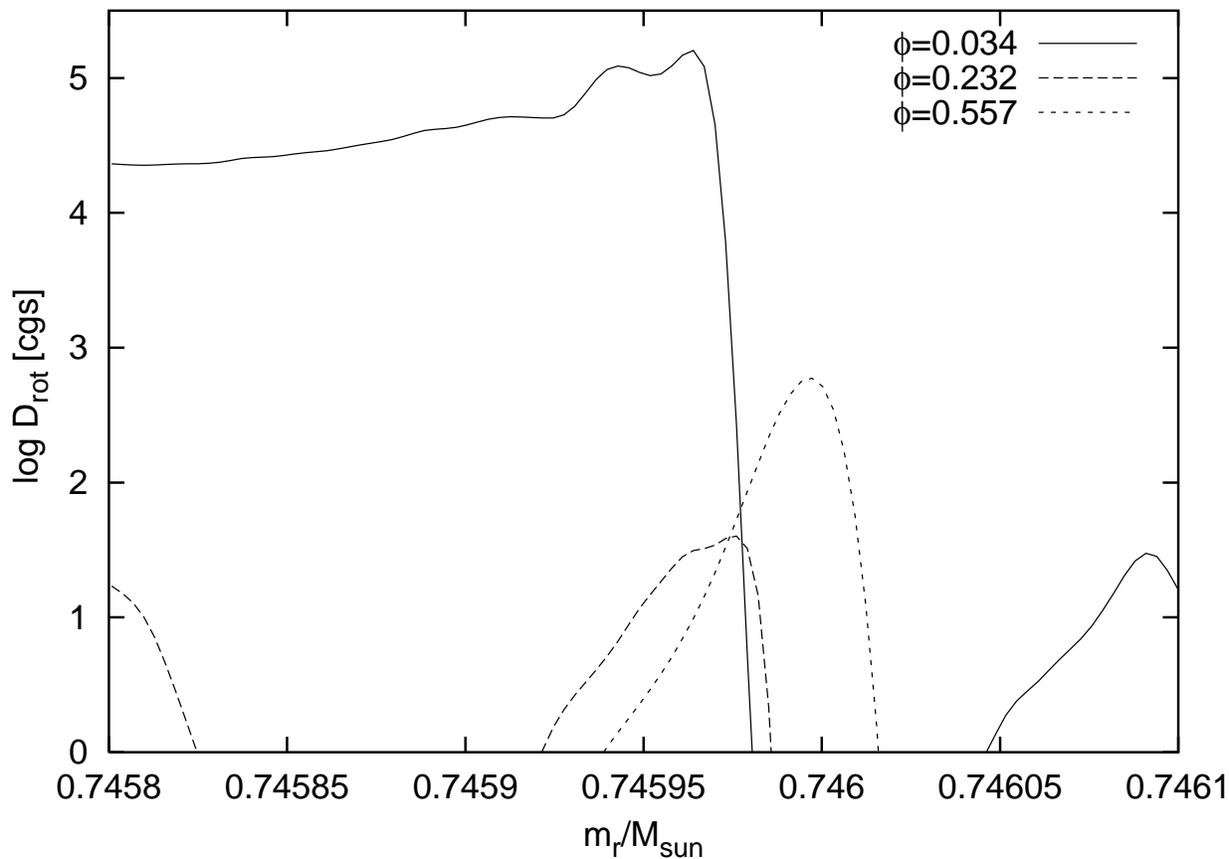} 
\figcaption{ \label{fig:Drot} 
Evolution of mixing coefficient due to rotationally induced mixing during the 
interpulse phase; 
labels give interpulse phase $\phi=(t-t_0)/\Delta t$, 
where the previous pulse, i.e. the maximum 
He-burning luminosity,
has occured at  $t_0$;  the interpulse period
is $\Delta t=3\cdot 10^{4}\jahre$.  
Two different mixing periods resulting from rotation occur in the partial mixing 
zone during the interpulse period (see text). 
}
\end{figure}

\clearpage
\newpage
\label{figcap}
Fig.\ \ref{fig:OV-prof} -- Abundance profiles in the partial mixing zone at three
times during the 7$^{\rm th}$ interpulse phase of the model with
  hydrodynamic overshoot and no rotation. Top panel: first post-processing model
after
  the end of the third dredge-up phase; middle panel: 
  $10\%$ of \cdr\ is burned by $\cdr(\alpha,\n)\ose$; bottom panel:
  neutron exposure $\tau$ at end of the \spr\ in the partial mixing
zone when the \cdr\ neutron source is exhausted. 

\begin{figure}
\plotone{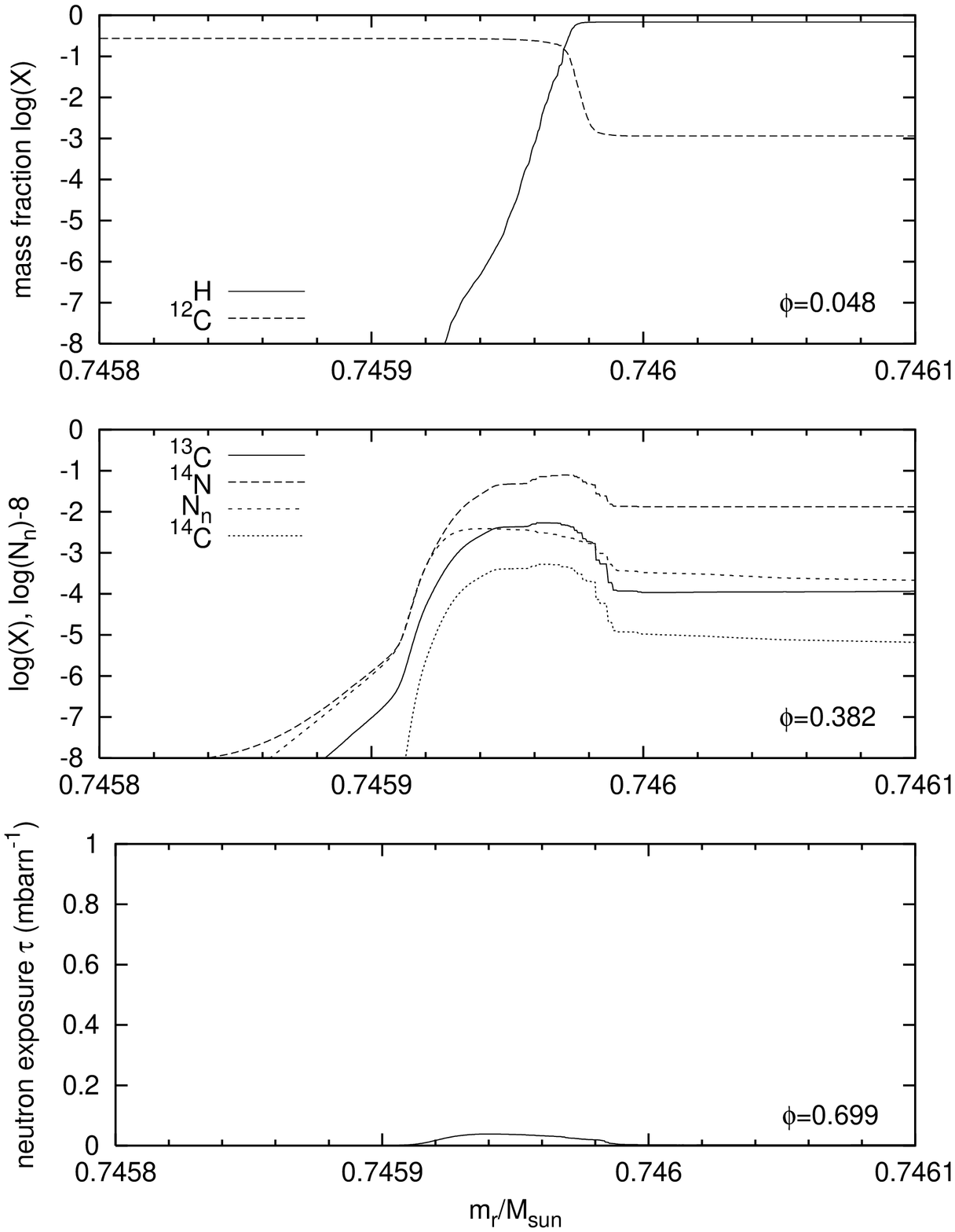} 
\figcaption{ \label{fig:ROT-prof} for caption see page \pageref{figcap}
}\end{figure}

Fig.\ \ref{fig:ROT-prof} -- Same as \abb{fig:OV-prof} for the 
simulation at the 25$^{\rm th}$ interpulse phase including rotation and no
hydrodynamic overshoot. 
The middle panel shows the same interpulse phase
as the middle panel in \abb{fig:OV-prof} with respect to the nuclear
lifetime of \cdr\ against $\alpha$-capture.
\newpage

\clearpage
\begin{figure}
\plotone{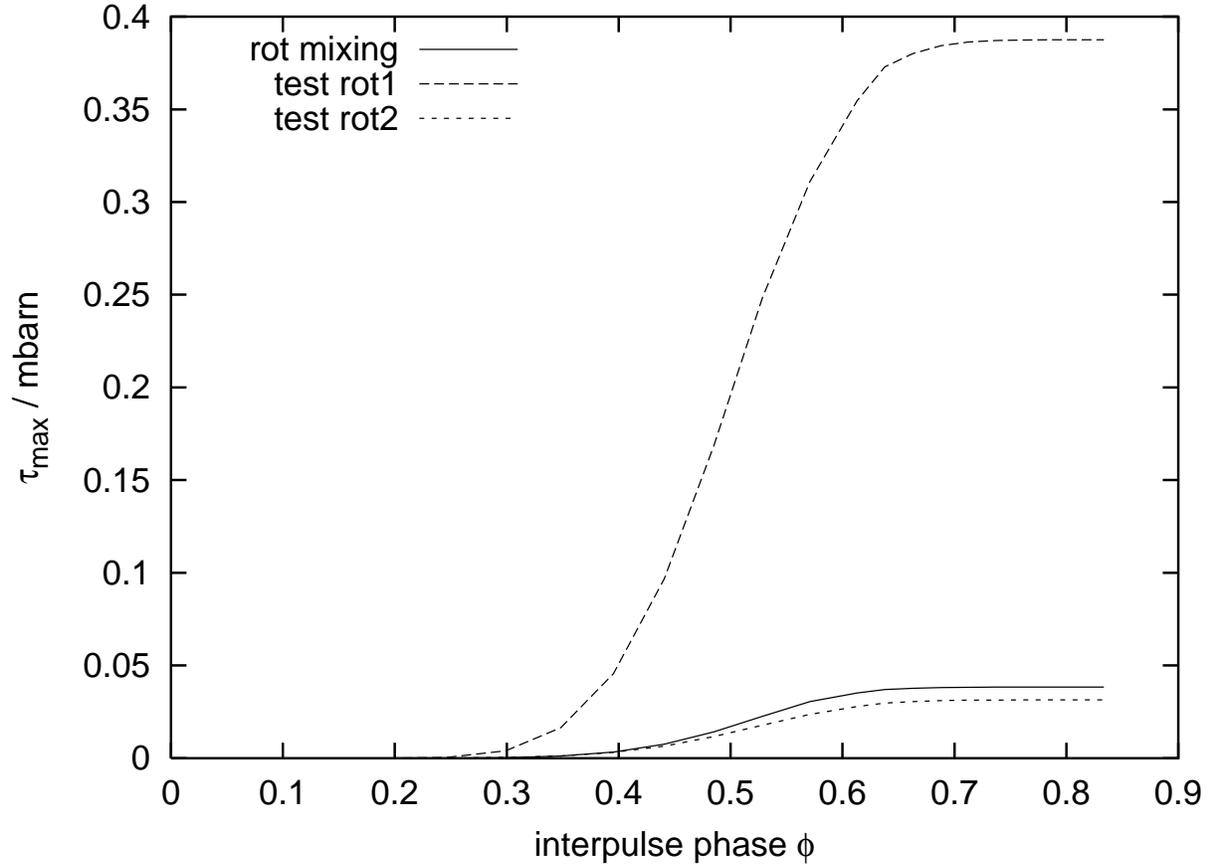} 
\figcaption{ \label{fig:taupeak} 
Maximum neutron exposure in the stellar layer where the \spr\ occurs during the
TP-AGB interpulse period as a function of phase $\phi$. 
The solid line is the result of detailed computation with rotational
mixing (shown in \abb{fig:ROT-prof}) and the long- and short-dashed lines show
test
calculations
(see text, \kap{sec:artmix}).
}\end{figure}
\begin{figure}
\plotone{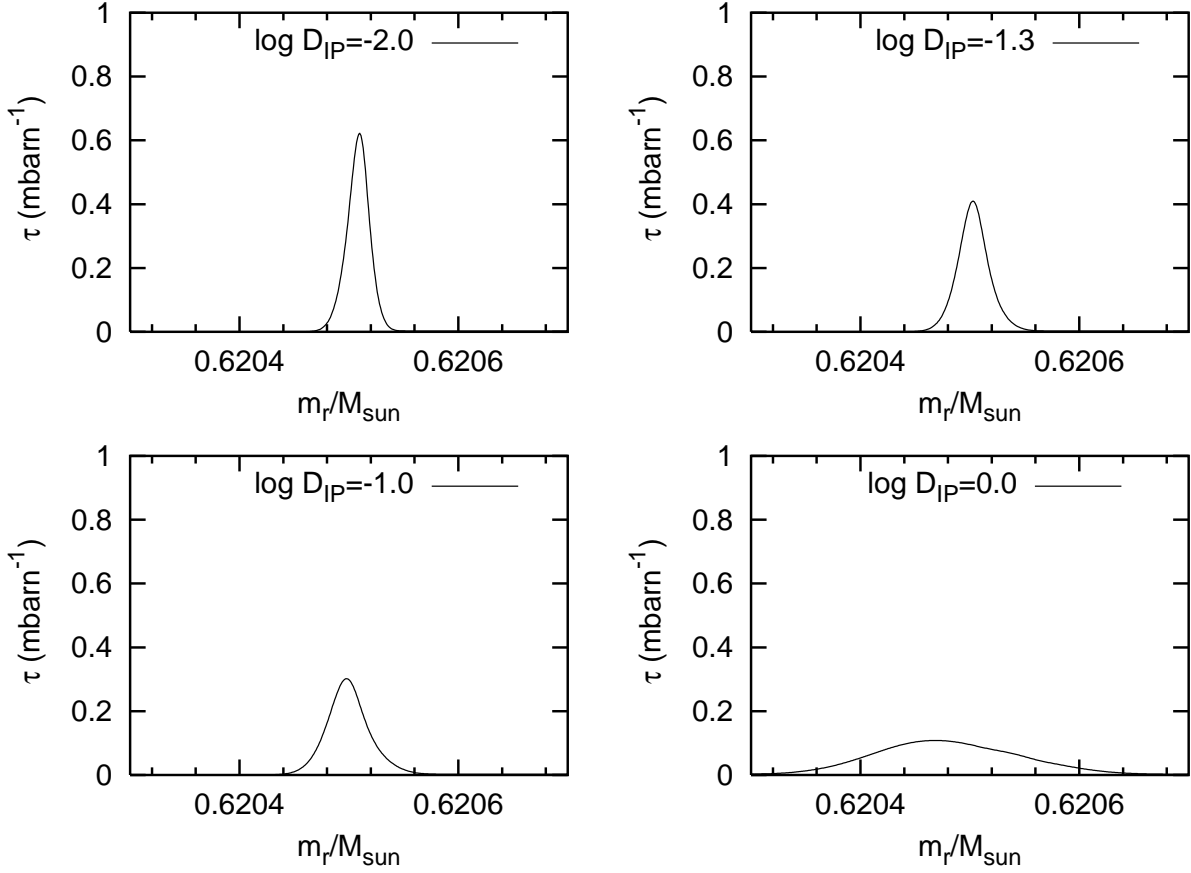} 
\figcaption{ \label{fig:OVROT-tau} 
Final neutron exposure profiles in the \spr\ layer of simulations
with a synthetic mixing law, combining the initial formation of a
partial mixing layer by overshoot with a constant interpulse mixing
coefficient that mimics the effect of shear mixing (see text for details).
}\end{figure}
\begin{figure}
\plotone{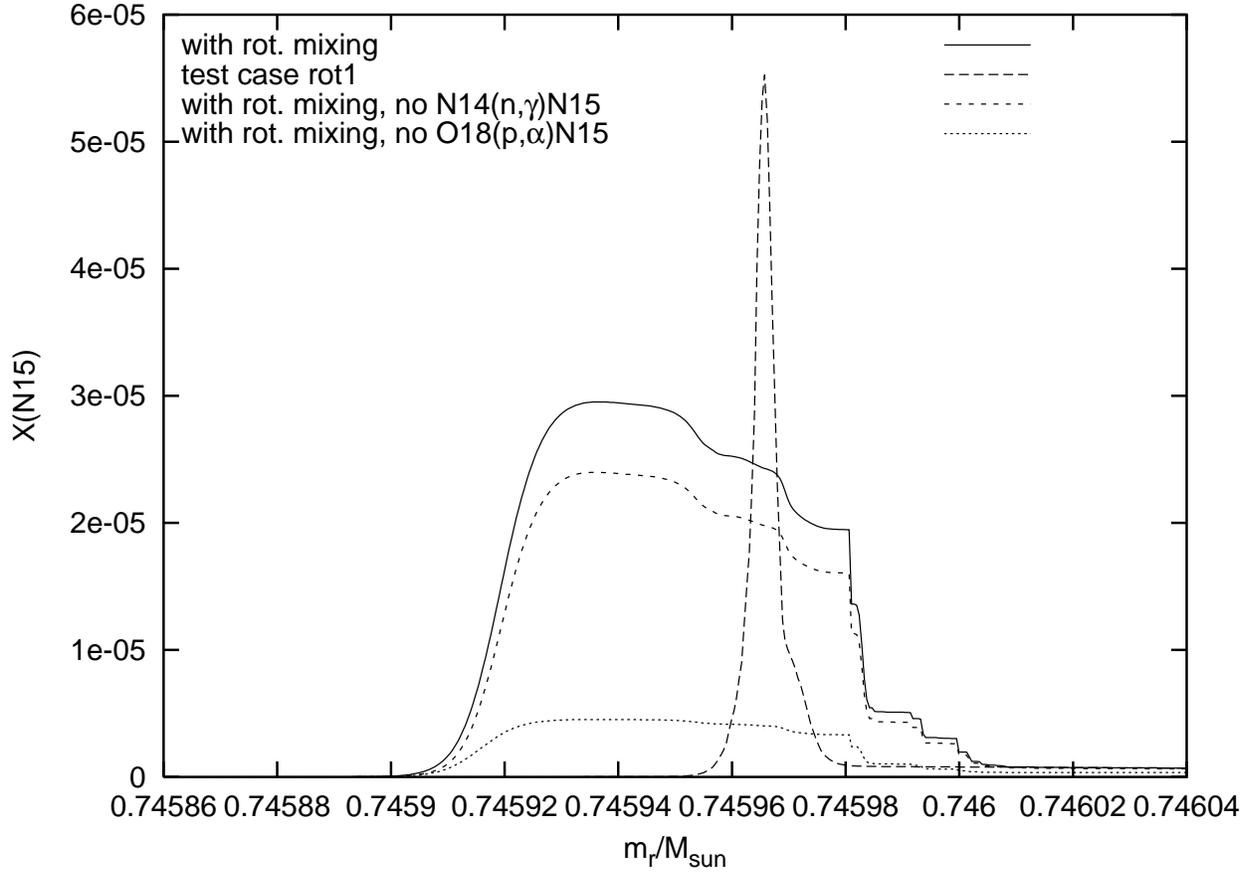} 
\figcaption{ \label{fig:N15} 
Profile of \nfu\ mass fraction across the partial mixing zone  at
a time close to the maximum neutron density in the simulation with
rotation presented in \kap{sec:rotmix}. the long-dashed and dotted
lines clarify the contributions by two reaction chains discussed in
the text. For comparison the \nfu\ profile for the same time is shown
for the test case rot1 presented in \kap{sec:artmix}.
}\end{figure}
\begin{figure}
\plotone{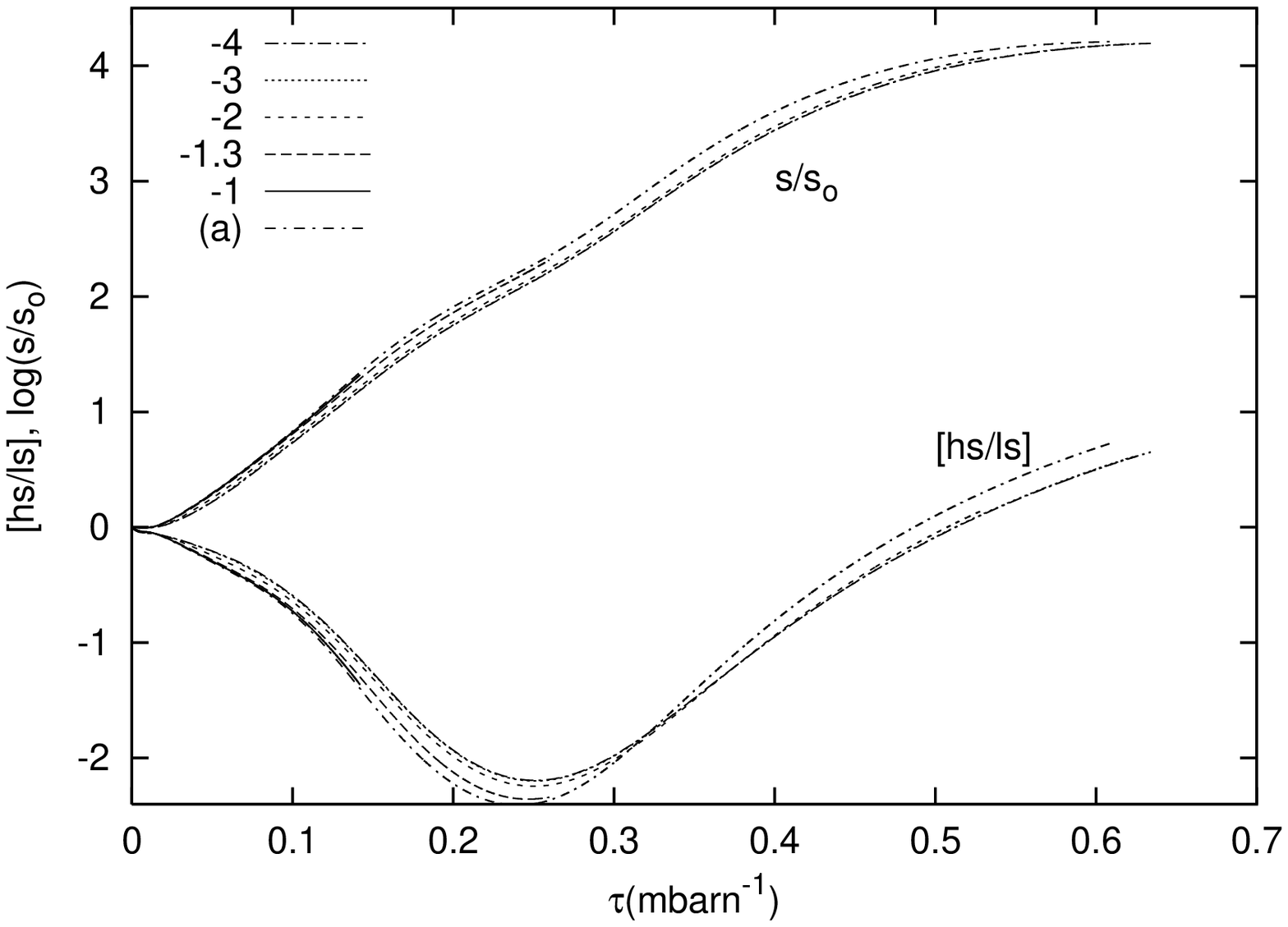} 
\figcaption{ \label{fig:tau-N14} 
Same as \abb{fig:tau-shsls}, but each case with a different initial
abundance of \nvi. The labels indicate the logarithm of the \nvi\ mass
fraction. Case (a) has been computed with intital abundances
$X(\hevi)=0.95$, $X(\czw)=0.03$, $X(\nvi)=1.2\cdot 10^{-5}$,  and
$\rho=20\mem{g/cm^3}$. } 
\end{figure} 

\end{document}